\documentclass[draftclsnofoot, 12pt, onecolumn]{IEEEtran}

\usepackage[english]{babel}
\usepackage{amsmath,amssymb}
\usepackage{times}
\usepackage{relsize}
\usepackage{graphicx}
\usepackage{url}
\usepackage{booktabs}
\usepackage{multirow}
\usepackage{mparhack}
\usepackage{subfigure}
\usepackage{authblk}
\usepackage{amsthm}
\usepackage{bm}
\usepackage{algorithm}
\usepackage[noend]{algorithmic}
\usepackage{array}
\usepackage{cite}
\usepackage{eqparbox}
\usepackage{mdwmath}
\usepackage{stfloats}
\usepackage{epsfig}
\usepackage{epstopdf}
\usepackage{xcolor}
\usepackage{pdfsync}
\usepackage{slashbox}
\usepackage{tikz}

\newtheorem{theorem}{Theorem}
\newtheorem{lemma}[theorem]{Lemma}

\newtheorem{corollary}[theorem]{Corollary}
\newtheorem{definition}{Definition}

\newcommand{\vertex}{\node[vertex]}

\newenvironment{remark}[1][Remark]{\begin{trivlist}
\item[\hskip \labelsep {\bfseries #1}:]}{\end{trivlist}}

\newcommand\T{\rule{0pt}{2.4ex}}
\newcommand\B{\rule[-1.0ex]{0pt}{0pt}}

\DeclareFontFamily{OT1}{pzc}{}
\DeclareFontShape{OT1}{pzc}{m}{it}%
              {<-> s * [1.1] pzcmi7t}{}
\DeclareMathAlphabet{\mathpzc}{OT1}{pzc}%
                                 {m}{it}

\def\CM{{\mathcal M}}
\def\CN{{\mathcal N}}
\def\CE{{\mathcal E}}
\def\CS{{\mathcal S}}
\def\CI{{\mathcal I}}
\def\CJ{{\mathcal J}}
\def\CR{{\mathcal R}}
\def\CC{{\mathcal C}}
\def\CL{{\mathcal L}}

\def\CV{{\mathcal V}}

\newcommand{\Cv}{\mathpzc{v}}

\def\PNAME{{CCS}}
\def\ALG{{AOCCS}}
\def\NEAR{{NCSA}}
\def\BOUND{{LEBS}}

\title{Optimal Cell Clustering and Activation for Energy Saving in Load-Coupled Wireless Networks}

\author[1]{Lei Lei}
\author[1,3]{Di Yuan}
\author[2]{Chin Keong Ho}
\author[2]{Sumei Sun}

\affil[1]{{\small Department of Science and Technology, Link{\"o}ping University, Sweden}}
\affil[2]{{\small Institute for Infocomm Research (I$^2$R), A$^*$STAR, Singapore}}
\affil[3]{{\small Institute for Systems Research, University of Maryland, College Park, MD 20740, USA}}
\affil[ ]{\em{{\small Emails: \{lei.lei@liu.se\},  \{diyua@itn.liu.se, diyuan@umd.edu\}, \{hock; sunsm\}@i2r.a-star.edu.sg}}}

\begin{document}

\date{}

\maketitle

\begin{abstract}
Optimizing activation and deactivation of base station transmissions
provides an instrument for improving energy efficiency in cellular
networks. In this paper, we study the problem of performing cell
clustering and setting the activation time of each cluster, with the
objective of minimizing the sum energy, subject to a time constraint
of serving the users' traffic demand.  
Our optimization framework accounts for inter-cell interference, and,
thus, the users' achievable rates depend on cluster formation.  We
provide mathematical formulations and analysis, and prove the
problem's NP hardness.  For problem solution, we first apply an
optimization method that successively augments the set of variables
under consideration, with the capability of approaching global
optimum. Then, we derive a second solution algorithm to deal with the
the trade-off between optimality and the combinatorial nature of
cluster formation.  Numerical results demonstrate that our solutions
achieve more than 40\% energy saving over existing schemes, and that
the solutions we obtain are within a few percent of deviation from
global optimum.
\end{abstract}

\vspace{-2mm}

\section{Introduction}
\label{sec:introduction}

Energy efficiency has become a major concern for cellular networks
due to the explosive growth of data traffic. Among the system
elements, base stations (BSs) account for more than $80\%$ of the total
energy consumption \cite{ict}, calling for new approaches for BS
operation. To this end, one solution is to coordinate and optimize the
activities of BSs, and the paradigm of BSs
operation has been shifted from ``always on'' to ``always available''
\cite{iccsleep}. Some underutilized BSs with low traffic can be turned off, for
example, to reduce the energy consumption, if the data traffic of the
BSs can be offloaded to other BSs.  Another
related scheme for energy saving is to organize the BSs by clusters
such that one cluster is active at a time. The cells within a cluster
are in transmission if and only if the cluster is active.
In this paper, we optimize cell cluster
formation and the activation time duration of each cluster,
with energy as the performance metric.

\vspace{-1mm}
\subsection{Related Works}

There are a number of studies that consider energy saving by
deactivating BSs \cite{offpeak1,offpeak2,offpeak3}.  In these works,
the periodic nature of cell's traffic, both temporally and spatially,
is exploited.  Energy consumption is reduced by deactivating some BSs
when the traffic demand is low.  If a BS is deactivated, its service
coverage is taken care of by other neighboring BSs that remain
active. Coordinated Multi-Point (CoMP) transmission can be applied,
see e.g., \cite{zoom}, to avoid coverage holes.

Energy saving can also be gained by deactivating BSs' power
amplifiers (PAs) if the amount of traffic does not require fully
continuous transmission.  In the transmission mode, the
PAs are accounted for most of the energy consumption.  Typically,
50-80\% of the total energy of a BS is consumed by the PAs \cite{ict}.
For long term evolution (LTE) systems, deactivating the PAs can be
done by adopting discontinuous transmission (DTX) at the BSs,
implemented by the use of Almost Blank Subframe (ABS) \cite{r10}.  In
\cite{dtx}, performance evaluation of DTX is carried out for a
realistic traffic scenario.

In BS scheduling, the BSs are grouped into clusters that potentially can
overlap, such that one
cluster is active (i.e., used for transmission) at a time, and a
schedule is designed to optimize the use of clusters to serve the user
demand with minimum energy.  In \cite{iccsleep}, the authors
assessed the performance of coordinated scheduling of BS
activation. In this case, inter-BS coordination is carried out for
groups of three cells, with pre-defined and fixed deactivation period
of each BS.  In \cite{conap}, the authors proposed a coordinated
activation scheme, in which the BSs are split into multiple BS groups. For
each group, the BSs switch between activation and deactivation
according to a pre-defined pattern. Simulation results in \cite{conap}
show that the scheme leads to 40\% less energy consumption.
In \cite{rayliu}, the authors
considered four BS deactivation patterns, to allow for progressively
deactivating BSs to improve energy efficiency, while maintaining the
quality of service (QoS).  Energy saving is achieved by dynamically
selecting the four patterns adaptively depending on the traffic
demand.

Another related topic is transmission scheduling in
wireless ad hoc and mesh networks (see, e.g., \cite{sinr,capone},
and the references therein). The task is to
organize links into groups, and determine the number of time
slots assigned to each group, in order to meet the demand with
minimum time (a.k.a. minimum-length scheduling). A subset of links can
form a group if and only if the signal-to-interference-and-noise ratio
(SINR) at the receivers meets a given threshold. A problem
generalization to continuous rates is studied in
\cite{tit}. In \cite{inria}, the authors studied 
transmission scheduling in mesh networks with a performance metric
that weights together time and energy.

\subsection{Our Work}

Most of the previous works for coordinated BS activation focus on
saving energy enabled by scenarios with relatively low user demand. For
the more general scenario with no specific assumption on user demand
level, energy-optimal BS scheduling for delivering the demand
within a strict time limit is challenging, due to the fact that the
achievable transmission rates within each cell are constrained by the
inter-cell interference. For LTE networks, the transmission rates
(i.e., demand delivered per time unit) in different cells are
inherently coupled with each other due to mutual interference.  To
characterize the achievable rates, we adopt the coupling model
in \cite{di,HoYuanSun13,icc2014,fullload} for cell load-dependent SINR.
Here, cell load refers to the utilization level of the time-spectrum
resource units (RUs) in orthogonal frequency division multiple access
(OFDMA). The cell load levels are coupled, i.e., they influence each
other. Namely, because the load reflects the amount of use of RUs for
transmission, the inter-cell interference generated by a cell to
another cell depends on the load of the former, and the interference,
in its turn, has impact on the load level of the latter.
In the load-coupling model, the dependency relation of the cell load
levels is taken into account in the SINR computation.
To the best of our knowledge, energy-efficient BS clustering
and scheduling, subject to maximum delay and rate characterization
based on the coupling relation among cells, has not been investigated
in the literature. 

In this paper, we formulate, analyze, and solve energy-efficient cell
clustering and scheduling (\PNAME{}), where the cells are required to
serve a target amount of data for the users within a time limit to
maintain an appropriate level of QoS, while considering the coupling
relation among cells due to interference.  Each cluster is a subset of
cells that are in simultaneous transmission mode, when the cluster is
active. Instead of pre-defined clusters, in \PNAME{} cell clustering
as well as cluster activation times are optimized. Within a cell, the
achievable rate vectors for the cell's users, taking into account
inter-cell interference, is not unique but form a rate region. Thus
solving
\PNAME{} also involves the selection of rate vectors.

We present the following contributions.  First, we formulate \PNAME{}
and prove its NP-hardness. A problem is called non-deterministic
polynomial-time hard, or NP-hard in short, if it is at least as
difficult as a large class of computational problems referred to as NP,
and, thus far, no polynomial-time algorithms exist for NP-hard
problems. As the next contribution, we present and prove a theoretical
result to enable to confine the consideration of rate vectors to a
finite set without loss of optimality.  On the algorithmic side, we
show how column generation
\cite{lp,BjVaYu03,capone} facilitates problem solving, and thereby
derive an algorithm for optimal cell clustering and scheduling
(\ALG{}) to approach the global optimum. Column generation is an
optimization method, in which a mathematical model is successively
expanded with new variables, such that the objective function gets
improved after each expansion, until the global optimum is reached. 
By our complexity results of
computational intractability, for large networks solving \PNAME{}
optimally is challenging.  
We then introduce our notion of locally enumerating interference, that is, 
for each BS, the rate evaluation of its users considers a selected small set of nearby
BSs as sources of interference, utilizing the fact that interference
from distant BSs is insignificant.
Using this notion, we present a
local-enumeration-based bounding scheme (\BOUND{}), providing lower
and upper bounds on the global optimum of minimum energy, as well as
enabling to deal with the trade-off between optimality and the
combinatorial nature of cluster formation. The bounds, in turn, serve
the purpose of gauging the deviation from optimality. Moreover, from
\BOUND{}, we derive a near-optimal cluster scheduling approach
(\NEAR{}). We present numerical results to illustrate the performance
of the proposed approaches. The results show significant energy
savings by \ALG{}, and the near optimality of solutions enabled by
\BOUND{} and \NEAR{}. We remark that, even though regular,
hexagon-shaped cells are used for performance evaluation for the
purpose of comparative study, our system model and the optimization
approaches do not impose any topological assumption, and hence they
are generally applicable to any given cellular network layout.

The rest of the paper is organized as follows.  Section II gives the
system model. In Section III, we formulate \PNAME{} and prove its
complexity.  Section IV presents algorithm \ALG{}. Section V details
the \BOUND{} scheme and \NEAR{}. Numerical results are given in
Section VI. Section VII concludes the paper.

{\it Notations}: We denote a (tall) vector by a bold lower case
letter, say $\bm{a}$, a matrix by a bold capital letter, say
$\bm{A}$. A set is denoted by a letter in calligraphic style,
say ${\mathcal A}$. Notation $\prec$ and $\preceq$ are for
componentwise inequalities between vectors.

\section{System Model}
\label{sec:system}

\subsection{Cellular Network with Cell Coupling}
\label{sec:network}

Consider a downlink OFDMA based cellular network with $I$ BSs serving
$J$ users.  We use $\CI=\{1, \dots, I\}$ and $\CJ = \{1, \dots, J\}$
to denote the sets of BSs and users, respectively. The set of users of
BS $i$ is denoted by $\CJ_i$, and user sets of all BSs form a
partitioning of $\CJ$.  Let $J_i = |\CJ_i|$, we have $\sum_{i \in \CI}
J_i = J$. Throughout the paper, we refer to BS $i$
interchangeably with cell $i$. In OFDMA, the time-frequency domain
resource is divided into resource units (RUs). A cell serves its users
by orthogonal (i.e., non-overlapping) use of the RUs. We use
$d_{ij}$ to denote the traffic demand (in bits) of user $j$ in cell
$i$. As a QoS requirement, all users' demands have to be served within
time $T$.

In the load-coupling model, the SINR computation over one RU uses the
cell load levels to take into account inter-cell interference. In the
following, we derive the SINR of one RU for user $j$ of BS $i$. We
denote by $p_i$ the transmission power per RU of cell $i$, and
$g_{ij}$ the channel gain. The noise effect is denoted by $\eta$,
which equals the power spectral density of white Gaussian noise times
the bandwidth of a RU.  For inter-cell interference from another BS
$k$ ($k
\not=i$), we use $p_k$ and $g_{kj}$ to denote the corresponding
transmission power and channel gain with respect to user $j$. Note
that interference is zero if BS $k$ is not utilizing any resource.
Following \cite{di,HoYuanSun13,icc2014,fullload}, we use the resource
utilization level of BS $k$ as a scaling factor in interference
modeling. With the given notation and discussion, the SINR of user $j$
in cell $i$ is formulated below.

\begin{equation}
\label{eq:sinr}
{\text{SINR}}_{ij}= \frac{p_{i}g_{ij}}{\sum_{k \in \CI \backslash \{i\}} p_k g_{kj} l_k+\eta}
\end{equation} 

In \eqref{eq:sinr}, entity $l_k$ is referred to as cell load, and
denotes the utilization level of RUs in cell $k$, that is, the
proportion of RUs allocated for transmission. The load vector is
denoted by ${\bm l} = [l_1, \dots, l_i, \dots, l_I]^T$. In
\cite{fullload}, it is shown that utilizing resource fully, i.e.,
${\bm l} = {\bm 1}$ is optimal from an energy standpoint.  However,
operating at full load means there is no spare OFDMA resource
units. For the sake of generality, our system model is formulated for
any preferred load level, with ${\bm 0} \prec {\bm l} \preceq {\bm
1}$. Note that in \eqref{eq:sinr}, the product $p_k g_{kj} l_k$
represents the amount of the interference from cell $k$ to user
$j$. The interference is Gaussian distributed in the worst
case. Therefore, by using Gaussian code, the achievable rate, in bits
per second, for user $j$ on one RU with bandwidth $B$ is computed as
$B \log_2(1+\mathrm{SINR}_{ij})$, where $B$ is the RU bandwidth.  Therefore, to
deliver a rate of $r_{ij}$ to user $j$ of cell $i$, $\frac{r_{ij}}{B
\log_{2} (1+{\text{SINR}}_{ij})}$ RUs are required.
Let $W$ denote the total number of RUs per cell.  
The corresponding load, i.e.,
the proportion of the RU consumption of cell $i$ due
to serving user $j$, is thus $l_{ij} = \frac{r_{ij}}{WB
\log_{2} (1+{\text{SINR}}_{ij})}$.
Observing that $l_{i}=\sum_{j \in \mathcal{J}_i}l_{ij} $ for cell $i$ gives the following
equation.

\vspace{-3mm}
\begin{equation}
\label {eq:rate}
l_{i}=\sum_{j \in \CJ_i} \frac {r_{ij}}{ WB\log_{2}
(1+\frac{p_{i}g_{ij}}{\sum_{k \in \CI \backslash \{i\}} p_k g_{kj}
l_k+\eta} )}, \ \forall i \in \CI
\end{equation} 

Without loss of generality, for convenience we normalize such that
$WB=1$. From \eqref{eq:rate}, one can observe that 
the users' rates cannot be set independently from each other.
Moreover, to satisfy the QoS
requirement, the rate values have to be chosen such that the demand is
delivered within time $T$ for all the users, that is, $T r_{ij} \geq
d_{ij}, \forall j \in \CJ_i, \forall i \in \CI$.

\subsection{Multi-Cell Clustering }

In Section \ref{sec:network}, we have given the basic elements of the system model
assuming that all cells are in transmission mode.
This may very well be feasible in meeting the QoS requirement, i.e.,
one can find rates for \eqref{eq:rate} such that all demands are
delivered within time $T$.  The strategy, however, may not be energy-optimal.
We now consider multi-cell clustering for energy optimization.
A \emph{cluster} refers to a subset of $\CI$, such that the BSs in the
subset are either all activated or all deactivated.  For
all possible $2^I-1$ non-empty subsets of $\CI$, denote by $\CS$ the index set:
$\CS = \{1, \dots, 2^I-1 \}$.  
Each index $s \in \CS$ maps to a unique subset of BSs.
Let $\CI_s$ denote the corresponding
set of cells of element $s \in \CS$.  Scheduling cluster $s$ means
that all the BSs in set $\CI_s$ are activated to be in transmission
mode to serve their associated users, whereas all the BSs in $\CI
\setminus \CI_s$ are deactivated.
In the latter case, the BS radio components are turned off and no data
can be transmitted.  There is a transition time between activation and
deactivation modes \cite{dtx}.  The transition time is however much
smaller than the entire scheduling period \cite{r10}, and hence we
consider the transition time to be zero in this paper.

Consider a cluster $s$ with BS set $\CI_s$. 
The equation \eqref{eq:rate} for cell $i \in \CI_s$ takes the following form.

\vspace{-2mm}
\begin{equation}
\label{eq:fullc}
{l_i=\sum_{j  \in \CJ_i}   \frac {r_{ij}^s}{ \log_{2} (1+\frac{p_{i}g_{ij}}{\sum_{k \in \CI_s \backslash \{i\}} p_kg_{kj}l_k +\eta}  )},  \hspace{0.2cm} \forall i \in \CI_s, ~\forall s \in \CS}
\end{equation} 

Here, $r_{ij}^s$ represents the rate allocated to user $j$ in cell $i$
within cluster $s$. As $l_i$ represents a preferred
resource utilization level of BS $i$, in this paper it is set
independently of cell clustering. 
Note that in \eqref{eq:fullc}, the user rates are
the variables.  By inspecting \eqref{eq:fullc}, we observe that it
forms a linear equation system of the user rates.  We
introduce the following entity.

\vspace{-3mm}
\begin{equation}
\label{eq:b}
          \begin{aligned}
          b_{ij}^s = \frac {1}{ \log_{2} (1+\frac{p_{i}g_{ij}}{\sum_{k \in \CI_s \backslash \{i\}} p_kg_{kj}l_k +\eta} )}, \forall j \in \CJ_i, \forall i \in \CI_s,  
\forall s \in \CS    
          \end{aligned}
\end{equation}

Then \eqref{eq:fullc} is simplified to the equation below.

\vspace{-3mm}
\begin{equation}
\label{eq:brl}
l_i = \sum_{j \in \CJ_i} b_{ij}^s r_{ij}^s, ~~~ ~ ~\forall i \in \CI_s, \forall s \in \CS  
\end{equation} 

For each cell $i$, its users are served when cell $i$ is
active. Thus, as we assume there is at least one user per cell, every
cell must be activated at least once, or, to be precise, every cell
must be included in at least one cluster that has positive activation
time. Note that a cell may be in multiple and active clusters. For
these clusters, the achieved rates of the cell's users and the
time durations of the clusters together determine the amount of served
traffic, which must meet the individual demand requirement within the
specified time limit.

We would like to point out that the system model focuses on downlink.
To support the downlink, some control traffic is necessary in the
uplink. This can be implemented by using time division duplex (TDD) or
frequency division duplex (FDD), as defined in 3GPP.

\begin{remark}
\label{re:onerate}
For any cell $i \in \CI_s$, there are infinitely many rate allocations satisfying 
\eqref{eq:brl}. Thus one can choose to activate a cluster multiple
times but with different rate allocations. In our system model, only
one rate allocation is to be selected for each cluster. However, as
will be clear later on, this seemingly strong restriction does not
impose any loss of generality. $\Box$
\end{remark}

\section{The Energy Minimization Problem}
\label{sec:problem}

\subsection{Problem Formulation}
\label{sec:firstformulation}

Energy-efficient \PNAME{} consists of determining the clusters that
shall be activated and the respective activation durations, and
the optimal user rate allocation within each cluster, such that the
sum energy is minimum and the users' demand are met within the time
limit.
For power consumption, we adopt a model that has been widely used 
(e.g., \cite{dtx,powermodel,Arnoldmodel}).  The power
of an active BS $i$ equals $p_i^{tot}=p_0 + l_iWp_i$.  The first component $p_0$
is load-independent to account for the auxiliary power consumption due
to processing circuits and cooling.  The second component represents
the transmission power with respect to the resource usage of BS $i$.
For an inactive BS, the power consumption is considered
negligibly small and assumed to be zero.  Thus the power
consumption of cluster $s$ is $p_s=\sum_{i \in \CI_s} p_i^{tot}$.
In the following we formally define the variables and formulate
the \PNAME{} problem.

In $P1$, the objective function \eqref{eq:p1obj} expresses
the sum energy, by taking the product of the sum power of each cluster
and its scheduled time duration. The QoS constraints
\eqref{eq:p1demand} and \eqref{eq:p1time} are imposed to ensure that the
required demand is delivered within the time limit.  Note that $i$ is
not a running index in the left-hand side of \eqref{eq:p1demand}. The
use of $i$ in the subscript of the summation is to exclude clusters
that do not contain cell $i$. Here, ``:'' means ``such that'' in an
optimization problem formulation.  Equations \eqref{eq:p1rate} define
the rate region.

$$
\begin{array}{ll}
\vspace{-1.2cm}
\\[1em] x_s & = \textrm{The time duration of activating the BSs in cluster $s$}.
\\ [1em] r_{ij}^s & = \textrm{The rate allocated to user $j$ of cell $i \in \CI_s$, $s \in \CS$.} 
\end{array}
$$

\vspace{-2mm}
\begin{subequations}
\label{eq:p1}
\begin{align}
P1:~~ \min~~ & \sum_{s \in \CS} p_s x_s \label{eq:p1obj}\\
\text{s.~t.} &  \sum_{s \in \CS: i \in \CI_s} x_s r_{ij}^s  \geq d_{ij},
 ~~\forall j \in \CJ_i, ~\forall i \in \CI \label{eq:p1demand} \\
& \sum_{s \in \CS} x_s  \leq T \label{eq:p1time} \\
& \sum_{j \in \CJ_i} b_{ij}^s r_{ij}^s = l_i, ~~~\forall i \in \CI_s, ~\forall s \in \CS  \label{eq:p1rate} \\
& x_s \geq 0, ~~\forall s \in \CS
\end{align}
\end{subequations}

We collect the user rate variables $r_{ij}^s$ and their coefficients
$b_{ij}^s$ of cell $i$ in cluster $s$ as column vectors ${\bm r}_i^s$ and
${\bm b}_i^s$, respectively. Then \eqref{eq:p1rate} has the following
compact form.

\vspace{-3mm}
\begin{equation}
\label{eq:simplex}
({{\bm b}_i^s})^T {\bm r}_i^{s} = l_i, ~\forall i \in \CI_s,~\forall s \in \CS  
\end{equation} 

We note that \eqref{eq:simplex} defines a simplex, which is a special
type of $J_i$-dimensional polytope, as the rate region of users of
cell $i$ in cluster $s$.  
Any point of this polytope represents an achievable rate vector, and vice versa.
We use $\CR_i^s$ to denote the simplex for cell $i \in \CI_s$ in cluster $s$.

Formulation $P1$ is non-linear and non-convex, due to
the product in \eqref{eq:p1demand}.  From the discussion above, in
general there are infinitely many possible rate vectors. However, we
will show this non-linearity can be overcome without loss of
optimality.

\begin{remark}
From \eqref{eq:p1}, the cell clustering problem is more
general than BS partitioning.  At optimum of \PNAME{},  
a cell may be in multiple active clusters
with different time durations. $\Box$
\end{remark}

\subsection{Linear Formulation of \PNAME{}}

Our first result is provided in Lemma \ref{th:finite} and Theorem
\ref{th:linear}. The result enables $P1$ to be transformed to
a linear but equivalent form with 
a finite number of rate allocations. 

\begin{lemma}
\label{th:finite}
Any solution to Problem $P1$ can be equivalently represented
using a finite number of rate vectors. 
\end{lemma}  
\begin{IEEEproof}
For any cluster $s$ and cell $i \in \CI_s$, the simplex, denoted by
$\CR_i^s$, is defined in \eqref{eq:simplex}.  Without loss of
generality, suppose the user indices of an arbitrary cell $i \in
\CI_s$ is $1, \dots, J_i$, and 
$\CI_s$ = $\{1, \dots, |\CI_s|\}$. Simplex $\CR_i^s$ has exactly
$J_i$ vertices ${\bm r}_i^{s,1}, \dots, {\bm r}_i^{s, J_i}$,
where ${\bm r}_i^{s,j}$ is the column vector having $\frac{l_i}{b_{ij}^s}$ as
its $j$th element and zero for all the other $J_i-1$
elements. Because $\CR_i^s$ is a convex set, any vector ${\bm r}_i^s
\in \CR_i^s$ can be represented as a convex combination of ${\bm
r}_i^{s,1}, \dots, {\bm r}_i^{s,J_i}$, that is, there exist
scalars $\theta_j \geq 0, \ j=1, \dots, J_i$, such that ${\bm r}_i^s =
\theta_1 {\bm r}_i^{s,1} + \theta_2 {\bm r}_i^{s,2} +
\dots + \theta_{J_i} {\bm r}_i^{s,J_i}$, and $\sum_{j=1}^{J_i}
\theta_j = 1$.

Suppose cluster $s$ is activated with time duration $x_s$ and rate
vectors ${\bm r}_i^s$, $i \in \CI_s$. For cell $i$, the vector
of the amount of served user demand is
given by multiplying scalar $x_s$ with the rate vector of this cell,
i.e., $x_s {\bm r}_i^s$. By the observation above, 
${\bm r}_i^s = \sum_{j \in \CJ_i} \theta_j {\bm r}_i^{s, j}$.
Hence, $x_s {\bm r}_i^s = \sum_{j \in \CJ_i} x_s \theta_j {\bm r}_i^{s, j}
= x_s\theta_1[\underbrace{\frac{l_i}{b_{i1}^s}, 0,\dots, 0}_{{\bm r}_i^{s,1}}]^T
+ \dots + x_s\theta_{J_i} [\underbrace{0,\dots, 0,
\frac{l_i}{b_{i J_i}^s}}_{{\bm r}_i^{s,J_i}}]^T$.
By this substitution, $x_s {\bm r}_i^s$ is equivalently expressed by a
weighted sum of rate vectors, each of which has one non-zero rate value.

For cluster $s$, denote by ${\bm r}^s$ the column vector obtained by stacking
${\bm r}_1^{s}, \dots, {\bm r}_{|\CI_s|}^{s}$, i.e., ${\bm r}^s =
[({\bm r}_1^{s})^T, \dots, ({\bm r}_{|\CI_s|}^{s})^T]^T$.  Activating
cluster $s$ with time duration $x_s$ (which is a scalar), the amount
of served user demand of the cluster, in vector form, is $x_s{\bm
r}^s=[x_s({\bm r}_1^s)^T, \dots, x_s({\bm r}_i^s)^T, \dots, x_s({\bm
r}_{|\CI_s|}^s)^T]^T$.  Applying the substitution step ${\bm r}_i^s =
\sum_{j \in \CJ_i} \theta_j {\bm r}_i^{s, j}$, and observing that
$\sum_{j \in \CJ_i} \theta_j = 1$, we obtain $x_s{\bm r}^s =
\theta_1[x_s ({\bm r}_1^s)^T, \dots, x_s({\bm r}_i^{s,1})^T, \dots, x_s ({\bm
r}_{|\CI_s|}^s)^T]^T+$ $, \dots, +\theta_{J_i}[x_s ({\bm r}_1^s)^T,
\dots, x_s ({\bm r}_i^{s,J_i})^T, \dots, \\ x_s ({\bm r}_{|\CI_s|}^s)^T]^T$.
Then, repeating the substitution procedure for the other cells leads
to the conclusion that the effect of activating cluster $s$ with any
rate vector ${\bm r}^s$ can be equivalently achieved by combining at most
$\Pi_{i \in \CI_s} J_i$ different rate vectors, and the lemma
follows.
\end{IEEEproof}

\begin{remark}
Lemma \ref{th:finite} further sheds light on 
the remark of Section \ref{sec:firstformulation}.  Consider a solution
in which a cluster is activated multiple times with different rate
allocations. Because each of them is equivalent to a
combination of the rate vectors from the same finite set, the
activations can be aggregated into one activation, for
which the rate allocation is derived from the coefficients used in the
combinations. Therefore considering one rate allocation per cluster in
problem formulation $P1$ does not cause any loss of generality. $\Box$
\end{remark}

\vspace{-3mm}
\begin{equation}      
\bm A^s=\left[                
  \begin{array}{ccccccccc}   
  \frac{l_i}{b_{i1}^s} & \frac{l_i}{b_{i1}^s} & \frac{l_i}{b_{i1}^s} &  0  &  0&  0&  0&  0&  0\\  
    0 & 0 &  0 &   \frac{l_i}{b_{i2}^s} &  \frac{l_i}{b_{i2}^s} &   \frac{l_i}{b_{i2}^s}&  0&  0&  0 \\
    0 & 0 &  0 &  0 & 0 &  0&  \frac{l_i}{b_{i3}^s}&  \frac{l_i}{b_{i3}^s}&  \frac{l_i}{b_{i3}^s} \\
   \frac{l_k}{b_{k4}^s} & 0 &  0 &  \frac{l_k}{b_{k4}^s} & 0  &  0&  \frac{l_k}{b_{k4}^s}&  0&  0\\  
   0 & \frac{l_k}{b_{k5}^s} &  0 &  0 & \frac{l_k}{b_{k5}^s} &  0&  0&  \frac{l_k}{b_{k5}^s}&  0 \\
    0 & 0 &  \frac{l_k}{b_{k6}^s} &  0 &0  &  \frac{l_k}{b_{k6}^s}&  0&  0&  \frac{l_k}{b_{k6}^s}\\
  \end{array}
\right]              
\end{equation}

Let ${\bm v}_i^s$ denote the set of vertices of
$\CR_i^s$. Collecting one element of each ${\bm v}_i^s$, $i \in
\CI_s$, leads to a column vector representing a rate allocation,
in which exactly one of the users in every cell has positive rate.
Enumerating all such combinations amounts to taking the Cartesian
product of sets ${\bm v}_i$, $\forall i \in \CI_s$.  This gives in
total $\Pi_{i \in \CI_s} J_i$ rate vectors, which we index by
$\CC_s=\{1, \dots, \Pi_{i \in \CI_s} J_i\}$.  As an example, consider
a cluster $s$ of two cells $\CI_s=\{i, k\}$ with three users in each
cell: $\CJ_i=\{1,2,3\}$ and $\CJ_k=\{4,5,6\}$.  The corresponding rate
vectors in $\CC_s$ can be expressed by a $\sum_{i \in \CI_s}
J_i$-by-$|\CC_s|$ matrix ${\bm A^s}$, where $\sum_{i \in \CI_s} J_i=6$
and $|\CC_s|=9$.

The vectors with index set $\CC_s$, i.e., the columns in $\bm A^s$ for
the example, are all feasible rate allocations for cluster $s$,
satisfying Equation (\ref{eq:p1rate}).  We denote the
rate allocated to user $j$ in $c \in \CC_s$ by $r_{ij}^{sc}$, $j \in
\CJ_i$, $i \in \CI_s$, and $c \in \CC_s$.
For each $i \in \CI_s$, there is one single user $j \in \CJ_i$ for
which $r_{ij}^{sc} = \frac{l_i}{b_{ij}^s}$, whereas the other users of the cell
have zero rates. For example, in the first column $
\underbrace{ [ \ \frac{l_i}{b_{i1}^s}, \ 0, \ 0}_{\text{cell} \ i}$, $
\underbrace{\frac{l_k}{b_{k4}^s}, \ 0, 0}_{\text{cell} \ k}$ $ ]^T$ of
$\bm A^s$, users 1 and 4 are allocated positive rates 
$r_{i1}^{s1}=\frac{l_i}{b_{i1}^s}$ and
$r_{k4}^{s1}=\frac{l_k}{b_{k4}^s}$ in the two cells, respectively.

We assign variable $x_{sc}$ for $c \in \CC_s$ to indicate the
activation time.  Next, we reformulate $P1$ as a linear formulation
$P2$, in which $x_{sc} \geq 0$ are variables,
whereas the rates are not.

\vspace{-1.5cm}
$$
\begin{array}{ll}
\\[1em] x_{sc} = \textrm{Activation time of cluster $s$ with rate index $c \in \CC_s$.} 
\end{array}
$$

\vspace{-2mm}
\begin{subequations}
\label{eq:p2}
\begin{align}
P2:~~ \min &  
  \sum_{s \in \CS}\sum_{c \in \CC_s} p_s x_{sc} \label{eq:p2demand} \\
\text{s.~t.} &     
   \sum_{s \in \CS: i \in \CI_s}\sum_{c \in \CC_s} r_{ij}^{sc} x_{sc}  \geq d_{ij}, \forall j \in \CJ_i, \forall i \in \CI \label{eq:p2time} \\
 &  \sum_{s \in \CS}\sum_{c \in \CC_s} x_{sc}  \leq T \\
 & x_{sc} \geq 0, \forall c \in \CC_s, \forall s \in \CS
\end{align}
\end{subequations}

The constraints in $P2$ have the same meaning as the first two
inequalities in $P1$. As $P2$ is restricted to a given and finite set
of rate vectors, the formulation is linear.

Recall that in $P1$, user rate $r_{ij}^s$ is an optimization variable,
and, for each cell in a cluster, the users' rates are subject to
\eqref{eq:p1rate} which defines the rate region that is a simplex. In
$P2$, $r_{ij}^{sc}$ is a not a variable. Specifically, $r_{ij}^{sc}, j
\in \CJ_i$, form a vector corresponding to a vertex of the simplex
defined by \eqref{eq:p1rate}. Utilizing the fact that any point of a
simplex can be equivalently represented by a convex combination of the
vertices of the simplex (cf.~Lemma~\ref{th:finite}), in $P2$ the
rate vectors representing the vertices are used instead of
\eqref{eq:p1rate}. Hence the $l$-parameters and $b$-parameters
do not appear explicitly in $P2$. Rather, they are used in calculating
the vertex vectors of the simplex. 

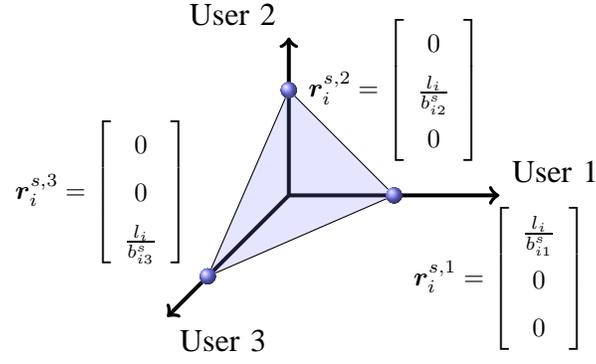
\begin{figure}[htbp]
\vspace{-5mm}
\tikzstyle{vertex}=[ball color=blue!50!,minimum size=6pt, inner sep=0.3pt,circle]
\[\begin{tikzpicture}[scale=0.7]
{\draw[color=black, line width=1.5pt,->] (0,0,0) -- (4,0,0) node[anchor=south west]{User 1};}
{\draw[color=black, line width=1.5pt,->] (0,0,0) -- (0,3,0) node[anchor=south east]{User 2};}
{\draw[color=black, line width=1.5pt,->] (0,0,0) -- (0,0,6) node[anchor=north west]{User 3};}
\vertex (r1) at (2,0,0) [label=-3: \small{${\bm r}_i^{s,1}=\left[ \begin{array}{c} \frac{l_i}{b_{i1}^s} \\ 0\\ 0\\ \end{array} \right]$}]{};
\vertex (r2) at (0,2,0) [label=right:\small{${\bm r}_i^{s,2}=\left[ \begin{array}{c} 0 \\ \frac{l_i}{b_{i2}^s}\\ 0\\ \end{array} \right]$}]{};
\vertex (r3) at (0,0,4) [label=170:\small{${\bm r}_i^{s,3}=\left[ \begin{array}{c} 0 \\ 0\\ \frac{l_i}{b_{i3}^s}\\ \end{array} \right]$}]{};
\path[fill=blue!50!, fill opacity=0.2] (2,0,0) -- (0,2,0)--(0,0,4)--cycle;
\path[fill=blue!50]
(r1) edge node[left]{} (r2)
(r2) edge node[left]{} (r3)
(r3) edge node[left]{} (r1);
\end{tikzpicture} \]
\vspace{-1.5cm}
\caption{An illustration: simplex $\CR_i^s$ and the vertices for three users.}
\label{fig:simplex}
\vspace{-0.7cm}
\end{figure}

It is instructive to illustrate Lemma \ref{th:finite} by an example.
Consider a single cell $i \in \CI_s$ serving three users
$\CJ_i=\{1,2,3\}$.  Figure~\ref{fig:simplex} provides an illustration
of the rate region defined by $b_{i1}^s r_{i1}^s + b_{i2}^s r_{i2}^s +
b_{i3}^s r_{i3}^s = l_i$.  This rate region corresponds to the surface
of the triangle. 
The three vertices are ${\bm r}_i^{s,1}=\left[\frac{l_i}{b_{i1}^s}, 0, 0
\right]^T$, ${\bm r}_i^{s,2}=\left[0, \frac{l_i}{b_{i2}^s}, 0 \right]^T$,
and ${\bm r}_i^{s,3}=\left[0, 0, \frac{l_i}{b_{i3}^s} \right]^T$. In $P1$,
the rate vector ${\bm r}_i^s$ has to be a point of the simplex, that is, 
$b_{i1}^s r_{i1}^s + b_{i2}^s r_{i2}^s +
b_{i3}^s r_{i3}^s = l_i$. Setting $\theta_j = \frac{r_{ij}^s b_{ij}^s}{l_i},
j=1, 2, 3$ gives $\theta_1+\theta_2+\theta_3 = 1$ and ${\bm r}_i^s =
\theta_1 {\bm r}_i^{s,1} + \theta_2 {\bm r}_i^{s,2} + \theta_3 {\bm r}_i^{s,3}$,
implying that ${\bm r}_i^s$ is a convex combination of the three vertices, which are
used in $P2$.

\begin{theorem}
\label{th:linear}
$P1$ and $P2$ are equivalent at optimum.
\end{theorem}  

\begin{IEEEproof}
From Lemma \ref{th:finite}, any solution of $P1$ can be equivalently
stated by a combination of a finite set of rate vectors. In addition,
from the construction of $P2$, the finite sets used in the proof of Lemma
\ref{th:finite} are exactly those in \eqref{eq:p2}. It then follows immediately
that any solution to $P1$ has an equivalent solution in $P2$.
Consider the opposite direction and take an arbitrary cluster
$s$ and its associated time durations $x_{sc}$, $\forall c \in \CC_s$,
in $P_2$.  For $\CC_s$, denote by ${\bm r}^{s1}, {\bm r}^{s2}, \dots,
{\bm r}^{s|\CC_s|}$ the corresponding rate vectors, all having length
$\sum_{i \in \CI_s} J_i$. 
We define rate vector ${\bm r}^s$ as
follows, where $x_{s} = \sum_{c \in \CC_s} x_{sc}$.

\vspace{-5mm}
\begin{equation}
\label{eq:newrate}
{\bm r}^s = \frac{x_{s1}}{x_s} {\bm r}^{s1} + 
\frac{x_{s2}}{x_s} {\bm r}^{s2} + \dots + \frac{x_{s|\CC_s|}}{x_s} {\bm r}^{s|\CC_s|}
\end{equation}

By construction in \eqref{eq:newrate}, ${\bm r}^s$ is a convex
combination of ${\bm r}^{s1}, {\bm r}^{s2}, \dots, {\bm
r}^{s|\CC_s|}$. Therefore for each cell $i \in \CI_s$, its
corresponding elements of ${\bm r}^s$ is in $\CR_i^s$, that is, ${\bm
r}^s$ is a feasible rate vector of cluster $s$ in $P1$.  Moreover, from
\eqref{eq:newrate}, it is evident that activating cluster $s$ with
time duration $x_s$ and rate vector ${\bm r}^s$ delivers exactly the
same amount of demand as activating ${\bm r}^{s1}, {\bm r}^{s2},
\dots, {\bm r}^{s|\CC_s|}$ with durations $\frac{x_{s1}}{x_s},
\frac{x_{s2}}{x_s}, \dots, \frac{x_{s|\CC_s|}}{x_s}$,
respectively. Hence any solution of $P2$ has an equivalent solution in
$P1$, and the theorem follows.
\end{IEEEproof}

\subsection{Problem Complexity}

Although $P2$ is linear, it is of exponential size in its complete
form, because there are $2^I-1$ candidate clusters. However, in
complexity theory, this fact, per se, does not prove problem hardness,
as a problem could be inappropriately stated in the formulation.
Therefore, in this section we formally conclude and prove
the hardness of \PNAME{}.

\begin{theorem}
\label{th:complexity}
\PNAME{} is NP-hard.
\end{theorem}
\begin{IEEEproof}
We give a polynomial-time reduction from the fractional chromatic
number in graphs \cite{LuYa94}. Consider a graph $G$ with $N$ nodes.
Denote by $\CV(G)$ the set of all independent sets of $G$, and $\CV(G,
n)$ the set of independent sets containing vertex $n$.  An independent
set is a set of non-adjacent nodes, i.e., no pair of the nodes in
the set is connected by an edge.  Each independent set $\Cv \in
\CV(G)$ is associated with a non-negative variable $x_\Cv$. Finding the
fractional chromatic number, which is NP-hard, amounts to $\{\min
\sum_{\Cv \in \CV(G)} x_\Cv; \text{s.t.} \sum_{\Cv \in \CV(G, n)} x_\Cv\geq
1, n=1,\dots,N \}$. The corresponding recognition version
is to determine if there is a solution with
$\sum_{\Cv \in \CV(G)} x_\Cv \leq K$ for a given number $K$. 
 
Consider the special case of \PNAME{} with $I = N$ BSs, each having a
single user. Thus we can use BS and user indices interchangeably.  Let
$\epsilon$ denote a positive number with $\epsilon \leq
2^{\frac{1}{N}}-1$.  For any BS $i \in \CI$, the parameters are as
follows: $p_i = 1$, $g_{ii}=\epsilon$, $l_i = 1$, and $d_{ii} =
1$. Moreover, $W=1$, $p_0=1$, and $\eta = \epsilon$.  For any two
BSs $i$ and $k$ with $i \not= k$, the channel gain $g_{ik} = 1$ if $i$
and $k$ are adjacent in graph $G$, otherwise $g_{ik}=0$.  The time
limit $T=K$.

We prove that at optimum of the defined \PNAME{}
instance, any two BSs connected by an edge in graph $G$ will not be in
the same cluster. Suppose the opposite, that is, at optimum there is
some cluster $s$ with time duration $x_s > 0$, and two BSs $i$ and $k$
that are adjacent vertices in $G$ are both present in $\CI_s$.
The cluster may contain additional BSs that are adjacent to $i$ or
$k$. Consider the subgraph composed by the nodes in $\CI_s$ and edges
between these nodes in graph $G$. Because $i$ and $k$ are adjacent,
there is a connected component in this subgraph containing $i$ and
$k$, possibly with additional BSs. Denote the nodes of this connected
component by $\CI_s(i,k)$. Suppose we combine $\CI_s \setminus
\CI_s(i,k)$ with each individual BS in $\CI_s(i,k)$. Doing so gives
$|\CI_s(i,k)|$ clusters, all with size $|\CI_s \setminus
\CI_s(i,k)|+1$. Consider activating these $|\CI_s(i,k)|$ new clusters,
each with time duration $\frac{x_s} {|\CI_s(i,k)|}$, in place of
cluster $s$.
For any BS in set $\CI_s \setminus \CI_s(i,k)$, the total time of
activation remains $x_s$, and the rate equals that of the BS in
$\CI_s$, because by the definition of $\CI_s(i,k)$, there is no
interference between the BSs in $\CI_s \setminus \CI_s(i,k)$ and those
in $\CI_s(i,k)$. 
For any BS in $\CI_s(i,k)$, the rate is strictly smaller than $\frac{1}{N}$ in cluster $s$ as $\CI_s(i,k)$ is a
connected component in graph $G$. For $i$, for example, the rate is no
more than
$\log_2(1+\frac{p_ig_{ii}}{p_{k}g_{ki}+\eta})=\log_2(1+\frac{\epsilon}{1+\epsilon})
< \log_2(1+2^{\frac{1}{N}}-1)=\frac{1}{N}$.  
Thus the demand delivered is less than $\frac{x_s}{N}$.  In the $|\CI_s(i,k)|$ new 
clusters defined above, the rate becomes 1, and hence with activation time $\frac{x_s}
{|\CI_s(i,k)|}$ the demand delivered becomes $\frac{x_s}
{|\CI_s(i,k)|}$, which is higher than $\frac{x_s}{N}$ as $|\CI_s(i,j)|
< N$.  Therefore, the amount of demand delivered via activating the
$|\CI_s(i,k)|$ clusters is no less than before.  Consider the energy
metric. For cluster $s$, the sum energy equals $(1+\epsilon) |\CI_s|
x_s$. For each of the new clusters, the sum power is
$(1+\epsilon)(|\CI_s \setminus
\CI_s(i,k)|+1)$. Because each is activated for time $\frac{x_s} {|\CI_s(i,k)|}$
and there are $|\CI_s(i,k)|$ clusters, the sum energy equals
$(1+\epsilon)(|\CI_s \setminus \CI_s(i,k)|+1)x_s$. This is smaller
than the sum energy of cluster $s$, because $|\CI_s
\setminus \CI_s(i,k)| \leq |\CI_s|-2$. Therefore, cluster $s$
cannot be optimal. In conclusion, at the optimum of the 
\PNAME{} instance, all clusters correspond to
independent sets in graph $G$. As $T=K$, solving the \PNAME{} instance
(or concluding its infeasibility) answers the
recognition version of fractional chromatic number. As the latter is NP-complete,
the theorem follows.
\end{IEEEproof}

\subsection{Two Simple BS Scheduling Strategies}
\label{sec:twosimple}
The previous analysis warrants the consideration of BS activation
strategies that are intentionally simplified for tractability.
Here we define two simple schemes: 1) individual activation of
each BS; 2) simultaneous activation of all BSs.

\begin{definition}
Using the notion of Time Division Multiple Access (TDMA),
a scheduling scheme is defined as ``TDMA" if 
one BS at a time is activated.
\end{definition}  

The TDMA scheme reduces the number of possible clusters from $2^I-1$ to
$I$, i.e., the total number of BSs. Utilizing Lemma \ref{th:finite},
one observes that with TDMA, it is optimal to serve one user at a time,
as formulated below.

\begin{lemma}
\label{th:uetdma}
For TDMA, then it is optimal for each BS to serve each
of its users individually, that is, TDMA at the BS level implies
time-division access of the users of each BS as well.  
\end{lemma}
\begin{IEEEproof}
From the proof of Lemma \ref{th:finite}, any achievable rate vector
${\bm r}_i^{TDMA}$ of BS $i$ under TDMA can be equivalently represented by
a combination of serving one user in $\CJ_i$ at a time.
Therefore, the TDMA scheme can be confined to deploying 
$J_i$ rate vectors, each having exactly one positive rate value
for one user in $\CJ_i$.
\end{IEEEproof}

From the lemma and \eqref{eq:fullc}--\eqref{eq:brl}, user $j \in
\CJ_i$ is served with the maximum possible rate $r_{ij}^{TDMA} = l_i \log_2(1 + \frac{p_i
g_{ij}}{\eta})$. Thus the time required for serving the user is
$t_{ij}^{TDMA} = \frac{d_{ij}}{l_i \log_2(1 + \frac{p_i
g_{ij}}{\eta})}$. The optimality
condition of TDMA is provided below.

\begin{theorem}
\label{th:tdma}
TDMA is optimal for \PNAME{} if it is feasible, i.e.,
if $\sum_{j \in \CJ_i} \sum_{i \in \CI} t_{ij}^{TDMA} \leq T$.
\end{theorem}  

\begin{IEEEproof}
Suppose at optimum of $P1$, a cluster $s$ of multiple BSs (i.e.,
$|\CI_s|>1$) is activated with time duration $x_s$, and denote by
${\bm r}_i^s$, $i \in \CI_s$ the rate vector allocated to BS $i$ in
the cluster. Consider replacing cluster $s$ with $|\CI_s|$ activations
of the individual BSs in $\CI_s$.  For any BS $i \in \CI_s$ with
single-BS activation, the corresponding rate vector ${\hat {\bm
r}}_i$ satisfies ${\hat {\bm r}}_i \succeq {\bm r}_i^s$, 
because there is no interference for single-BS activation and thus the ${\bm b}$ vector in
\eqref{eq:simplex} becomes smaller.
Therefore, if each single BS of the cluster is activated with time
$x_s$, $x_s{\hat {\bm r}}_i \succeq x_s {\bm r}_i$, i.e., the demand
that is served is no less than that of cluster $s$. Therefore, to
deliver the same amount of demand $x_s {\bm r}_i$ to the users in any
BS $i \in \CI_s$, the time required by single-BS activation of $i$,
denoted by ${\check x}_i$, satisfies ${\check x}_i \leq x_s$.  The
energy consumed by cluster $s$ equals $x_s
\sum_{i \in \CI_s} p_i^{tot}$.  With single-BS activations the
energy consumption is improved to $\sum_{i \in \CI_s} p_i^{tot}
{\check x}_i$.  The total time duration of the latter is $\sum_{i \in
\CI_s} {\check x}_i$, which however may be higher than $x_s$.  Hence,
as long as the time limit $T$ is not exceeded, replacing
cluster $s$ with single-BS activations improves energy, and the
theorem follows.
\end{IEEEproof}

By Theorem \ref{th:tdma}, TDMA is the preferred strategy for energy
efficiency if the users' traffic demand is low such that it can be
met by TDMA within the time limit. Thus in this paper, we are more
interested in scenarios of heavier traffic, for which TDMA is not
time-feasible, i.e., $\sum_{j \in \CJ_i} \sum_{i \in \CI}
t_{ij}^{TDMA} > T$.

In addition to TDMA, we consider, as a simple and baseline scheme,
the conventional strategy of having all BSs constantly
activated. This scheme, as defined below, will be used as a benchmark
for performance comparison.

\begin{definition}
A scheduling scheme is defined as ``All-on" if all the BSs are
constantly transmitting until all users' demands have been met.
\end{definition}  

In All-on, one cluster $s'$ containing all the BSs is activated.  Each
BS serves its users with relatively lower rates due to the
worst-case interference.  Denote by $t_1, t_2, \dots, t_J$ the
transmission times required for meeting the individual user demands.
The total activation time in All-on is a constant
$T_{all-on}=\max \
\{t_1, t_2, \dots, t_J\}$ which is the longest transmission time for
serving an individual user's demand.  If $T \geq T_{all-on}$, All-on
is feasible and the sum energy is $p_{s'}T_{all-on}$,
otherwise All-on is infeasible.  Note that the rate vectors to be used
are subject to optimization, and the algorithm in the next section,
i.e., Algorithm \ref{alg:cga}, can will used to obtain the optimal rates of ``All-on''
for performance comparison.  All-on in this paper is defined to be
consistent with \cite{rayliu} in order to enable a reasonable
comparison in Section \ref{sec:results}.

\section{Optimization Algorithm for Cell Clustering and Scheduling}
\label{sec:cg}

\subsection{Outline}

In this section, we propose and present an optimization algorithm for
optimal cell clustering and scheduling (\ALG{}). Consider formulation
$P2$. It is in linear form, though the number of clusters is
exponential in network size. However, most of the clusters are of no
significance for constructing the optimal solution.  
In fact, as formalized below, one can
conclude the existence of an optimal solution using at most $J+1$
clusters.

\begin{lemma}
\label{th:lpbasic}
For any feasible instance of \PNAME{}, there exists an optimal
solution activating at most $|J+1|$ clusters.
\end{lemma}  

\begin{IEEEproof}
By theory of linear programming (LP) \cite{lp}, if an LP formulation
is feasible and bounded, then at least one optimum is a so called
basic solution. The two conditions hold by the lemma's assumption and
the structure of $P2$, respectively. For any basic solution of $P2$,
the number of variables in the base matrix equals $J+1$, i.e., the
number of constraints.  At an optimal basic solution, therefore, the
number of $x$-variables with positive values does not exceed $J+1$,
and the lemma follows.
\end{IEEEproof}

In view of the size of $P2$ and Lemma \ref{th:lpbasic}, 
\PNAME{} should be solved in some other way than using 
$P2$ as is. Toward this end, we consider
a column generation \cite{cg} approach for solving
\PNAME{} with guaranteed global optimality.
The resulting algorithm \ALG{} is based on a decomposition of $P2$
into a master problem and a pricing problem.  The decomposition
procedure keeps a small subset of candidate clusters in the master
problem.  The solution quality of the master problem is then
successively improved by adding new clusters and rate vectors which
are generated from solving the pricing problem.

\subsection{The Master Problem}

The so called master problem is a restricted form of $P2$.
 A cluster along with an
associated rate vector of each cell in the cluster are jointly
represented as a ``column".  Adding a cluster and associated rates to
the master problem is then equivalent to generating a new column in
the coefficient matrix of $P2$.  The master problem is presented
below; the difference from $P2$ is that the complete sets of clusters
$\CS$ and rate vectors $\CC_s$ are replaced by subsets $\breve{\CS}$
and $\breve{\CC_s}$, respectively, that are successively augmented by
new columns.

\vspace{-0.7cm}
\begin{subequations}
\label{eq:p3}
\begin{align}
{P3:~~} \min~~ & \sum_{s \in \breve{\CS}}\sum_{c \in \breve{\CC_s}} p_s x_{sc} \label{eq:p3obj}\\
\text{s.~t.} &  \sum_{s \in \breve{\CS}: i \in \CI_s}\sum_{c \in \breve{\CC_s}} r_{ij}^{sc} x_{sc}  \geq d_{ij}
 ~~\forall j \in \CJ_i, \ \forall i \in \CI \label{eq:p3demand} \\
& \sum_{s \in \breve{\CS}} \sum_{c \in \breve{\CC_s}}x_{sc}  \leq T \label{eq:p3time} \\
& x_{sc} \geq 0, c \in \breve{\CC_s}, s \in \breve{\CS}
\end{align}
\end{subequations}

One iteration of \ALG{} amounts to solving the master problem
\eqref{eq:p3}, and determining if augmenting
\eqref{eq:p3} by a column (i.e., a cluster and an associated rate vector) 
that is not present in
\eqref{eq:p3} can improve \eqref{eq:p3obj}.
This is achieved by solving the pricing problem, to examine whether or
not there exists any new column with a negative reduced cost
\cite{cg}.

\subsection{The Pricing Problem}
\label{sec:pricing}

For the optimum of \eqref{eq:p3}, denote by $\pi_{ij}^*$ and
$\lambda^*$ the dual variable values associated with constraints
(\ref{eq:p3demand}) and (\ref{eq:p3time}), respectively. From linear
programming, the reduced cost of a given cluster $s$ and a candidate
rate vector $c \in \CC_s$ is equal to $p_s - \sum_{i \in \CI_s}\sum_{j \in \CJ_i}
r_{ij}^{sc} \pi_{ij}^*-\lambda^*$. 
Here, $r_{ij}^{sc}$ is a not a variable, because it is associated with
a given candidate rate vector $c \in \CC_s$.
Thus, finding the column with the minimum reduced cost can be
performed for one cluster at a time. For each cluster $s \in \CS
\setminus \breve{\CS}$, the task is to find the rate vector index $c \in
\CC_s$ for which the reduced cost attains its minimum for the given cluster.
Recall that the cardinality of $\CC_s$ is $\Pi_{i \in \CI_s}
J_i$, which can be very large. However, this task can be equivalently 
formulated by the following linear optimization formulation.

\vspace{-7mm}
\begin{subequations}
\label{eq:p4}
\begin{align}
{P4:~~} \omega_s = \max~~ & \sum_{i \in \CI_s} \sum_{j \in \CJ_i} \pi_{ij}^* r_{ij}^s \label{eq:p4obj}\\
\text{s.~t.} &  \sum_{j \in \CJ_i} b_{ij}^s r_{ij}^s = l_i, \forall i \in \CI_s \label{eq:p4simplex} \\
& r_{ij}^s \geq 0, \forall j \in \CJ_i, \forall i \in \CI_s \label{eq:p4nonnegative}
\end{align}
\end{subequations}

In formulation \eqref{eq:p4}, $r_{ij}^s, j \in \CJ_i, i
\in \CI_s$, are the optimization variables. Their values are
chosen to minimize the objective function \eqref{eq:p5obj} 
that represents reduced cost, subject to
\eqref{eq:p4simplex}-\eqref{eq:p4nonnegative} that
define the rate region.

\begin{remark}
As $P4$ is a linear program, the optimum is located at a vertex of the simplex
defined by \eqref{eq:p4simplex}. Thus the resulting rate vector indeed
qualifies for formulation $P2$, i.e., the rate vector, represented by optimization variables
$r_{ij}^s, j \in \CJ_i, i \in \CI_s$, is one of the elements in $\CC_s$ $\Box$
\end{remark}

After solving \eqref{eq:p4} for each cluster, if $\min_{s \in \CS
\setminus \breve{\CS}} p_s - \omega_s - \lambda^* < 0$, then the
corresponding cluster and its rate vector are added as a new column to
augment the master problem $\eqref{eq:p3}$. If the minimum is non-negative,
then the optimum of $P3$ with the current $\breve{\CS}$ is also the
global optimum for $P2$.  The \ALG{} operations are given in Algorithm
\ref{alg:cga}.

\begin{algorithm}[ht!]
\begin{algorithmic}[1]
\caption{\ALG{}}
\label{alg:cga}
\STATE Construct $P3$ with an initial set of clusters $\breve{\CS}$
\REPEAT 
\STATE Solve the master problem $P3$. \label{step:master}
\FOR{$s \in \CS \setminus \breve{\CS}$}
\STATE Solve the pricing problem $P4$ 
\ENDFOR
\IF{$\min_{s \in \CS \setminus \breve{\CS}} p_s - \omega_s - \lambda^* < 0$}
\STATE Add the corresponding cluster and rate vector to $\breve{\CS}$ and $\breve{\CC_s}$, respectively
\ENDIF
\UNTIL{$\min_{s \in \CS \setminus \breve{\CS}} p_s - \omega_s - \lambda^* \geq 0$} 
\end{algorithmic}
\end{algorithm}
\vspace{-8mm}

\begin{remark}
The global optimality of Algorithm~\ref{alg:cga} does not depend on
the specific choice of the initial subset $\check{\CS}$. For example,
$\check{\CS}$ could have only one cluster containing all the cells.
In Algorithm 1, $\check{\CS}$ and the rate vectors for each $s \in
\check{\CS}$ are successively augmented by new clusters and rate
vectors, such that the objective function value of $P3$ becomes
improved after each augmentation. Identifying which cluster and rate
vector to add is the task in the pricing problem. By 
linear programming theory \cite{lp}, solving the pricing problem
will either lead to a cluster and rate vector for augmenting
$P3$, or conclude none of
the remaining clusters and rate vectors has negative reduced cost. In
the latter case, global optimality is reached. $\Box$
\end{remark}

The computational bottleneck of \ALG{} is on the pricing problem $P4$.
Even if \eqref{eq:p4} is linear, to ensure global optimality
\eqref{eq:p4} needs to be solved for all clusters, and the number of
clusters is exponential in the network size. 
To this end, in the next section we develop an algorithm with a control
parameter for the trade-off between complexity and optimality.

Although Algorithm~\ref{alg:cga} is presented for static problem input,
the column generation approach has the potential of addressing system
dynamics in respect of the number of users and their demands. By
column generation, the elements of clusters and rate vectors are
successively added. When there is an update in the input, say changed
user demand, the algorithm simply starts from Step~\ref{step:master},
utilizing the current sets of clusters ${\check
\CS}$ and rate vectors ${\check \CC}_s, s \in {\check \CS}$,
i.e., a warm start, instead of optimizing by starting from scratch.  If there is
a new user, adding zero as the rate for this user in the current rate
vectors, along with the current scheduling solution at hand (which satisfies the
demands of all other users), together achieve the warm-start effect.

\section{Local Enumeration Based Bounding Scheme}

The challenge in dealing with the complexity of the pricing problem
lies in the coupling relation between cells. Specifically, the
interference and hence the rate region of one cell depend on the
cluster composition, and the number of possible clusters is exponential
in the number of BSs.

We introduce a concept that we refer to as local
enumeration. The notion is to confine, for each cell, the interference
consideration to its local neighborhood. This is motivated by the fact
that, for any BS, the interference experienced is dominated by the BSs
nearby, whereas interference coming from more remote BSs is
insignificant.  For a cluster and any of its cells, inter-cell
interference originates from all other cells in the cluster.  Suppose
we need to determine the cells to be grouped together to form a new
cluster in some optimization process (e.g., solving the pricing
problem in Section~\ref{sec:pricing}). For each candidate cell, there
are $2^{I-1}$ possible interference scenarios, depending on which of
the remaining $I-1$ cells are to be included in the same cluster. If
we only account for which cells nearby are included in the cluster in
interference calculation, the number of combinations of interference
scenarios to be considered becomes much smaller.  As will be clear
later on, the size of the local neighborhood acts as a control
parameter for the trade-off between the accuracy of interference
estimation and complexity reduction. Moreover, the solution scheme via
local enumeration allows to compute upper and lower bounds to the
global optimum of \PNAME{}, as well as a near-optimal BS clustering
and scheduling solution.

\subsection{Local Enumeration}
\label{le-a}

In local enumeration, the interference calculation of each cell is
restricted to a selected set of cells that are nearby. For cell $i$,
denote by $M_i$ the number of cells to be included in interference
consideration, with $1 \leq M_i \leq I-1$. The selection of the $M_i$
cells could be, for example, based on sorting the cells in $\CI
\setminus \{i\}$ using the average interference that each of them generates, 
if active, to the users in cell $i$. Denote by $\CL_i$ the resulting
set of cells after the selection. Then, enumeration of the
interference scenario for cell $i$ takes place for the $M_i$ cells in
$\CL_i$. That is, the enumeration applies to all possible combinations
of active cells in $\CL_i$, giving $2^{M_i}$ combinations in total,
including the case where no cell in $\CL_i$ is selected.  We denote $\CE_i$ as
the collection of all combinations of $\CL_i$ where each combination
is augmented with cell $i$.  In other words, only the interference
from the cells in $\CL_i$ are exactly accounted for.  Parameter $M_i$
controls the size of enumeration. Note that if $M_i = I-1$, then
all cells are part of interference consideration and the scheme
falls back to global enumeration.

\begin{figure}[htbp]
\vspace{-1cm}
\centering
\includegraphics[scale=0.26]{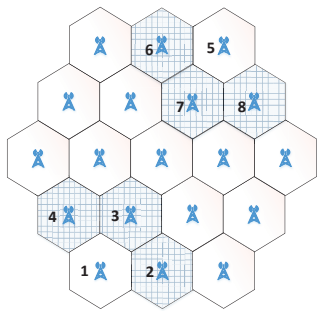}
\vspace{-1.2cm}
\caption{An illustration of local enumeration.}
\label{fig:le}
\end{figure}

An example is given in Figure~\ref{fig:le}.  Suppose $M_1 = M_5 = 3$,
meaning that interference from three cells will be considered for cell
1 and cell 5, respectively. The resulting cells for interference
consideration are $\CL_1 = \{2, 3, 4\}$ and $\CL_5 =
\{6,7,8\}$, respectively.
Local enumeration of the cells in $\CL_i$ and $\CL_5$ gives the
combinations shown in Table \ref{tab:letable}.

\vspace{-5mm}
\begin{table}[htbp]
\caption{Enumeration of $\CL_1$ and $\CL_5$ for cells $1$ and $5$ in Figure~\ref{fig:le}.}
\vspace{-7mm}
\label{tab:letable}
{\scriptsize
\centering
\begin{tabular}[t]{ll}
\hline
\bf{$\CE_1$:}  & $\{1\},\{1,2\},\{1,3\},\{1,4\},\{1,2,3\},\{1,2,4\},\{1,3,4\},\{1,2,3,4\}$ \\
\bf{$\CE_5$:}  & $\{5\},\{5,6\},\{5,7\},\{5,8\},\{5,6,7\},\{5,6,8\},\{5,7,8\},\{5,6,7,8\}$ \\
\hline 
\end{tabular}
}
\end{table}

To avoid potential notational conflict, we denote by $\CM_i = \{1,
\dots, 2^{M_i}\}$ the index set of $\CE_i$, and denote by $\CN_{ei}$
the set of cells associated with $e \in \CM_i$. For any $e
\in \CM_i$, the rate region of cell $i$ is defined, such that 
only the activations of cells in $\CN_{ei} \setminus
\{i\}$ are accounted for exactly.
To see the effect, consider as an example two clusters $s_1$ and
$s_2$, with $\CI_{s_1}=\{1,3,4,5\}$ and $\CI_{s_2}=\{1,3,4,6,7\}$.
For cell $1$, in both cases the corresponding element of $\CE_i$ in
local enumeration is $\{3,4\}$, i.e., the two significant interferers
in both clusters. Therefore, from cell $1$'s viewpoint, the cluster
solutions at the network level have a many-to-one mapping to the
elements in $\CN_{e1}$, leading to dramatically reduced complexity in
comparison to enumerating all the $2^{I}-1$ rate regions.

Recall that parameters $b_{ij}^s (j \in \CJ_i, i \in \CI_s, s \in
\CS)$ are the coefficients in equation \eqref{eq:p1rate} of
cell $i$ in cluster $s$. With local enumeration, the equation of
a cell $i$ is defined with respect to the cells in $\CL_i$.
To avoid any ambiguity in notation, we use $\beta_{ij}^e$ to denote
the corresponding parameter for user $j$ of cell $i$, for the
interference scenario $e \in \CM_i$.

We consider two options of treating the less significant interference
from cells outside $\CL_i, \forall i \in \CI$, corresponding to the
best and worst possible interference scenarios, respectively.  In the
first option, interference from the BSs in $\CI \setminus (\CL_i \cup
\{i\})$ is considered zero, no matter of whether they are in the same
cluster as cell $i$ or not.  Hence the interference is considered for
the cells in $\CL_i$ only, giving the following definition of the
$\beta$-parameter.

\vspace{-2mm}
\begin{equation}
\label{eq:leoff-b}
{\check \beta}_{ij}^e = \frac{1}{\log_{2} (1+\frac{p_{i}g_{ij}}{\sum_{k \in \CN_{ei} \setminus \{i\}} p_k g_{kj} l_k +\eta} )} 
\end{equation} 
\vspace{-2mm}

In the worst-case scenario, all BSs outside $\CL_i$ are considered
being active concurrently, irrespective of the true status.
The resulting parameter definition is given below.

\vspace{-2mm}
\begin{equation}
\label{eq:leon-b}
{\hat \beta}_{ij}^e = \frac{1}{ \log_{2} (1+\frac{p_{i}g_{ij}}{\sum_{k
\in (\CN_{ei} \setminus \{i\}) \cup (\CI \setminus (\CL_i \cup \{i\}))}
p_kg_{kj}l_k +\eta})}
\end{equation} 
\vspace{-2mm}

\subsection{Bounding Scheme \BOUND{}}

Based on local enumeration, we develop a scheme \BOUND{} to
provide lower and upper bounds to the global optimum.
In \BOUND{}, column generation is applied using the same master
problem as in Section \ref{sec:cg}, whereas the pricing problem is
re-formulated by using local enumeration of interference scenarios.
In $P5$, we present the variable definitions of the new
formulation for pricing, and then the formulation itself.

\vspace{2mm}
$z_{i}$ = 
    $\left\{
      \begin{array}{l l}      
        1 & $if cell $i$ is selected for cluster formation,$\\
        0 & $otherwise$. \\
      \end{array}
    \right.$

\vspace{2mm}
$y_{ei}$ = 
    $\left\{
      \begin{array}{l l}      
        1 & $if cluster formation corresponds to $e \in \CM_i$ for cell $i$, i.e., the active cells in$\\
        & $ $\CL_i \cup \{i\}$ are $\CN_{ei}$$, \\
        0 & $otherwise$. \\
      \end{array}
    \right.$

\vspace{2mm}

$r_{ij}^e = $ the rate of user $j \in \CJ_i$ for $e \in \CM_i$. 

\vspace{-5mm}
\begin{subequations}
\label{eq:p5}
\begin{align}
P5:~~\max~~ & 
   \sum_{i \in \CI} \sum_{e \in \CM_i} \sum_{j \in \CJ_i} 
   \pi_{ij}^* r_{ij}^e - \sum_{i \in \CI} p_i^{tot} z_i \label{eq:p5obj} \\    
\text{s.~t.}~~ & 
\sum_{j \in \CJ_i} \beta_{ij}^e r_{ij}^e = l_i z_i, ~\forall e \in \CM_i, \forall i \in \CI & \label{eq:p5rate}\\
 & \ \ \  \ \sum_{e \in \CM_i} y_{ei} = z_i,~\forall i \in \CI & \label{eq:p5choice} \\
 & \ \   \sum_{e \in \CM_i: h \in \CN_{ei}} y_{ei} \leq z_h, ~\forall h \in \CL_i, \forall i \in \CI & \label{eq:p5on}\\
 &   1 - \sum_{e \in \CM_i: h \in (\CL_i \cup \{i\}) \setminus \CN_{ei}} \hspace{-7mm} y_{ei} \geq z_h, ~\forall h \in \CL_i, \forall i \in \CI & \label{eq:p5off} \\
 & r_{ij}^e \geq 0, \forall j \in \CJ_i, \forall e \in \CM_i, \forall i \in \CI & \\  
 & y_{ei} \in \{0, 1\}, \forall e \in \CM_i, \forall i \in \CI & \\  
 & z_{i} \in \{0, 1\}, \forall i \in \CI & 
\end{align}
\end{subequations}

Similar to Section \ref{sec:pricing}, the objective \eqref{eq:p5obj}
is to minimize the reduced cost, or equivalently to maximize its
negation. The second term in \eqref{eq:p5obj} accounts for the total
cluster power. For cell $i$, \eqref{eq:p5rate} defines the rate
regions in the local enumeration of the interference scenarios, taking
into account whether or not cell $i$ is to be part of cluster
formation.  If cell $i$ is selected to be active, then exactly one of
the scenarios in cell $i$'s local enumeration of interference has to hold
true, otherwise none of the scenarios will apply.  These effects are
achieved by \eqref{eq:p5choice}. The next two sets of inequalities
state the relation between clustering at the network level and the
resulting interference scenarios of local enumeration.  Note that,
each of the interference scenarios of a cell $i$ implies which of the
cells in $\CL_i$ are active, and vice versa.  For example,
interference scenario $\{1,2,4\}$ of cell $1$ in Figure
\ref{fig:le} applies if and only if cells $2$ and $4$ are active (i.e., part
of the cluster formation) and cell $3$ is inactive.  In other words,
there must be consistency between the $z$-variables and $y$-variables.
This consistency is achieved by \eqref{eq:p5on}--\eqref{eq:p5off}.  By
\eqref{eq:p5on}, for any cell $i$ and another cell $h$ that is subject
to interference consideration, the latter must be active (i.e.,
$z_h=1$) if any of the $y$-variables corresponding to interference
scenarios containing $h$ is set to one. Consider again the
aforementioned example.  If the interference scenario $\{1,2\}$ is
selected for cell $1$, then $z_2$ must be one.  Inequalities
\eqref{eq:p5off} deliver a similar effect for the opposite case,
namely the choice of interference scenario of cell $i$ also dictates
the cells that must be inactive in $\CL_i$.

From a scalability point of view, the strength of $P5$ is that the
interference enumeration is limited to the cells in $\CM_i$, of which
the size is $2^{M_i}-1$ for each $i \in \CI$. This is in contrast to
the pricing problem in Section \ref{sec:pricing} for which the number
of candidate clusters is $2^I-1$. As $\CM_i$ contains
neighboring BSs with significant interferences only, typically $M_i \ll I$
without much loss of accuracy. Moreover, $M_i$ can be used as a 
control parameter for the trade-off between accuracy and computation.

\begin{remark}
At the optimum of $P5$, the cluster solution is given by cells for
which $z_i=1, \forall i \in \CI$. For each of such cells, there is an optimal
rate vector corresponding to a vertex of the simplex defined by
\eqref{eq:p5rate}, because the objective function 
is linear in rate. Thus the cluster and the rate vector obtained
from solving $P5$ are similar to the columns in $P2$, in the sense
that for any cell in the cluster, exactly one user will attain a
positive rate, and the other users have zero rates. $\Box$
\end{remark}

In solving \eqref{eq:p5}, the parameters $\beta_{ij}^e$ $(j \in \CJ_i,
i \in \CI_s, e \in \CM_i)$ are set to ${\check \beta}_{ij}^e$ or ${\hat
\beta}_{ij}^e$ in \eqref{eq:leoff-b} and \eqref{eq:leon-b}, corresponding
to treating the BSs outside the local enumeration (LE) scope $\CL_i$ to be 
all non-active and all active, respectively. We use ``LE-off'' and
``LE-on'' to respectively refer to the two settings. These settings,
when embedded into the column generation algorithm \ALG{}, yield
lower and upper bounds confining the global optimum.
This result is formalized below.

\begin{theorem}
\label{th:bound}
Denote by $E^*$ the global optimum of \PNAME{}, and $E^*_{\text{LE-off}}$
and $E^*_{\text{LE-on}}$ the optimal values from embedding 
LE-off and LE-on into column generation, respectively. Then
$E^*_{\text{LE-off}} \leq E^* \leq E^*_{\text{LE-on}}$.
\end{theorem}  

\begin{IEEEproof}
Denote by $\CS_{\text{LE-on}}^*$ the set of clusters in the optimal
solution from the LE-on scheme. For any cluster $s \in
\CS_{\text{LE-on}}^*$, the interference scenario in the local
enumeration for cell $i \in \CI_s$, induced by $s$, is the index
element $e \in \CM_i$ such that $\CN_{ei} = (\CL_i \cup \{i\}) \cap \CI_s$.  Denote
by $e_i(s)$ the index of this interference scenario.  From the remark
above the theorem, for each $s \in \CS_{\text{LE-on}}^*$ and cell $i
\in \CI_s$, exactly one user of $i$, say $j^*$, has positive rate
$r_{ij^*}^{e_i(s)} = \frac{l_i}{{\hat \beta}_{ij^*}^{e_i(s)}}$,
whereas all other users of $\CJ_i$ carry zero rates.

Consider replacing the rate of $j^*$ by $r_{ij^*}^s =
\frac{l_i}{b_{ij^*}^s} $, while keeping the zero rates of the other
users of cell $i$. 
By definition, $\CN_{e_i(s)i}  \subseteq \CI_s$ in LE-on. 
Therefore $\sum_{k \in (\CN_{e_i(s)i} \setminus \{i\}) \cup (\CI \setminus (\CL_i \cup \{i\}))} p_kg_{kj^*} \geq \sum_{k \in \CI_s
\setminus \{i\}} p_k g_{kj^*}$. 
From \eqref{eq:brl} and \eqref{eq:leon-b}, $b_{ij^*}^s \leq
\hat{\beta}_{ij^*}^{e_i(s)}$, and thus $r_{ij^*}^s \geq r_{ij^*}^{e_i(s)} 
$.  Performing this rate update for all cells in $\CI_s$, we obtain a
column $c \in \CC_s$ in $P2$ for cluster $s$, with a rate vector such
that the values are at least as high as those in the rate vector in
the solution of LE-on with the same cluster, and the non-zero elements
coincide in their positions.  Thus for the same time duration of each
$s \in \CS_{\text{LE-on}}^*$, deriving the corresponding columns of
$P2$ gives a feasible, though not necessarily optimal, solution of
$P2$. Hence $E^* \leq E^*_{\text{LE-on}}$.

For the second inequality, the idea of the proof is
analogous, though the starting point is the globally optimal set of
clusters of $P2$. The proof consists in observing that each cluster
and its associated rate vector correspond to a solution that is
potentially returned by solving $P5$, but with the same or higher rate
values; the latter is because for any cluster $s$, $i \in \CI_s$,
interference scenario $e_i(s)$, and $j \in \CJ_i$, we have
$\check{\beta}_{ij}^{e_i(s)} \leq b_{ij}^s$.  By
the theory of column generation in linear programming \cite{cg}, 
$\check{\beta}_{ij}^{e_i(s)} \leq b_{ij}^s$ implies
that the optimal value from LE-off will not under-perform $E^*$, hence
$E^*_{\text{LE-off}} \leq E^*$.
\end{IEEEproof}

\subsection{Near-Optimal Solution Based on \BOUND{}}

\BOUND{} not only provides bounds to the global optimum, but also 
enables the computation of a feasible solution of \PNAME{}. From the proof of
Theorem \ref{th:bound}, for LE-on, starting from
$\CS_{\text{LE-on}}^*$ and the rate allocation for each $s \in
\CS_{\text{LE-on}}^*$, and replacing each positive rate value with
that derived from \eqref{eq:brl} leads to a feasible solution of $P2$.
Note that the cardinality of $\CS_{\text{LE-on}}^*$ is at most $J+1$,
thus computing this feasible solution comes with little additional
effort. The idea leads to the following near-optimal cluster
scheduling approach (\NEAR{}).

\begin{enumerate}
\item $\breve{\CS} \leftarrow \CS_{\text{LE-on}}^*$.
\item If $r_{ij}^{e_i(s)} > 0$, $r_{ij}^s \leftarrow \frac{l_i}{b_{ij}^s}$, 
otherwise $r_{ij}^s \leftarrow 0$, $\forall j \in \CJ_i, \forall i \in \CI_s, \forall s \in \breve{\CS}$.
\item Solve $P3$ to optimality.
\end{enumerate}

Note that the rate values used in LE-on are pessimistic, i.e., they
are equal to or lower than the values derived from
\eqref{eq:brl}. Thus the total energy given by \NEAR{}, denoted 
by $E_{\NEAR{}}^*$, improves that of LE-on, giving the corollary below.

\begin{corollary}
\label{th:newbound}
$E^* \leq E_{\NEAR{}}^* \leq E_{LE-on}^*$.
\end{corollary}  

Note that a feasible solution may be derived from LE-off as well.
However, since in LE-off the rate values are on the optimistic side,
there is no guarantee that the scheduling time limit $T$ can be respected
after replacing the rate values with those obtained from
accurate interference calculation.

\section{Performance Evaluation}
\label{sec:results}

\subsection{Experimental Setup}

\begin{figure}[htbp]
\vspace{-11mm}
\centering
\includegraphics[scale=0.27]{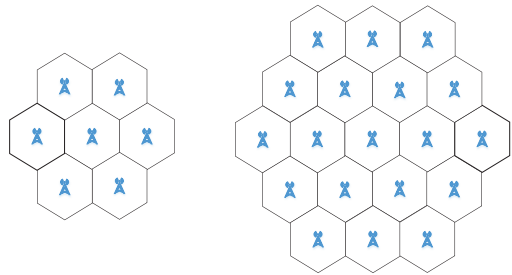}
\vspace{-1.6cm}
\caption{Networks used for performance evaluation.}
\label{fig:network}
\end{figure}

Two networks consisting of seven and nineteen cells, respectively,
have been used in the simulations, see Figure~\ref{fig:network}.  Each
BS serves five randomly and uniformly distributed users within the
cell's area.
The networks operate at 2~GHz.  Following the LTE
standards, we use one resource block to represent a resource unit with
180 kHz bandwidth in the simulation.  The total bandwidth amounts to
4.5~MHz.  The channel gain consists of path loss and shadowing
fading. The path loss follows the COST-231-HATA model.  For shadowing,
the log-normal distribution with 8~dB standard deviation is used. For
each network, we generate one hundred instances and consider the
average performance. Motivated by the results in
\cite{icc2014,fullload}, we set cell's load ${\bm l}={\bm 1}$. In 
Algorithm~\ref{alg:cga}, ${\check \CS}$ is initially set to contain
all clusters of size two, with ${\check \CC}_s = \CC_s$ for each $s
\in {\check \CS}$.  Table
\ref{tab:parameter} summarizes the key simulation parameters.

\vspace{-0.5cm}

\vspace{-2mm}
\begin{table}[htbp]
\caption{Simulation Parameters.}
\vspace{-1cm}
\label{tab:parameter}
\centering
\begin{tabular}[t]{l l}
\hline
\bf{Parameter}  &   \bf{Value}   \\
Cell radius  & 500~m \\
Carrier frequency  & 2~GHz \\
Total bandwidth per cell & 4.5~MHz \\
Bandwidth per RU & 180~kHz \\
Number of users per cell & 5 \\
User demand $d_{ij}$ & 2~Mbits \\
Path loss & COST-231-HATA\\
Shadowing & Log-normal, 8~dB standard deviation \\
Transmit power $p_i$ per RU  &  1~W   \\
Circuit power $p_0$ per BS  &  5~W   \\
Noise power spectral density & -174~dBm/Hz \\
Load per BS & 1.0 \\
\hline
\end{tabular}
\end{table}

Among the algorithms, \ALG{} guarantees global optimality (see also
the remark in Section~\ref{sec:pricing}), however it is not intended for large
networks.  Algorithm \NEAR{} is a sub-optimal algorithm providing a
heuristic solution, by means of local enumeration by which the pricing
problem is of polynomial size. The purpose of \BOUND{} is to deliver
bounds on global optimum (which is hard to compute for large
networks), and thereby enable to evaluate \NEAR{} in terms of
the deviation from global optimality. In the following,
we present and compare the results of these algorithms.

\subsection{Energy Optimization by \ALG{} and \NEAR{}}

To evaluate the performance of the proposed \ALG{} and \NEAR{},
the conventional scheme ``All-on" (see Section
\ref{sec:twosimple}) and a scheme called ``BS Switch-off Pattern
Strategy (BSPS)" proposed in
\cite{rayliu}, have been implemented for comparison.  
For BSPS, five activation patterns, referred to as All-on, I, II, III,
IV, are proposed in \cite{rayliu}.  The first pattern coincides with
our ``All-on" scheme defined in Section~\ref{sec:twosimple}. The other
four patterns are composed by cell subsets with decreasing
cardinality. In \cite{rayliu}, one of the patterns is
chosen at a time based on the level of user demand. 
We remark that inter-cell interference is not considered
for analytical simplicity in \cite{rayliu}. 
For our simulation, however, we account for inter-cell interference
in the comparison.
For the comparative study, we consider the best achievable
performance of BSPS, by allowing mixed and optimized use of its
patterns. This is carried out by generating cell clusters based on the
patterns in \cite{rayliu}, followed by solving the resulting
optimization formulation \eqref{eq:p2} to global optimality.

\vspace{-0.7cm}
\begin{table}[htbp] 
\addtolength{\tabcolsep}{-5pt}
\caption {The energy consumption comparison}
\vspace{-7mm}
\label{tab:res}
\centering
\begin{tabular}[t]{|l|c|c|c|c|c|c|}
\hline
\multirow{2}{*}{ \bf{7-Cell Network}}   &  \multicolumn{6}{c|}{ \bf{Energy Consumption (Joule) }} \\
\cline{2-7}
&  \multicolumn{1}{c|}{$T$=1 (s)} & \multicolumn{1}{c|}{$T$=1.5} & \multicolumn{1}{c|}{$T$=2} & \multicolumn{1}{c|}{$T$=2.5} & \multicolumn{1}{c|}{$T$=3} & \multicolumn{1}{c|}{$T$=3.5} \\
\hline    	  

\bf{\ALG{}}  &143.76  &133.81    &130.82  &129.62 & 129.24 & 129.05 \\

\bf{\NEAR{} ($M_i$=5)}   &144.09  &134.96   &131.84   &129.84   &129.26   &129.07\\
\bf{BSPS in \cite{rayliu}} &147.11  &140.45  &139.71  &139.25  &139.04 &139.01  \\
\bf{All-on} &221.32	&221.32	&221.32	&221.32 &221.32 &221.32 \\
\hline
\multirow{2}{*}{\bf{19-Cell Network}}   &  \multicolumn{6}{c|}{\bf{Energy Consumption (Joule)}} \\
\cline{2-7}
&   \multicolumn{1}{c|}{$T$=2 (s)} & \multicolumn{1}{c|}{$T$=2.5} & \multicolumn{1}{c|}{$T$=3} & \multicolumn{1}{c|}{$T$=3.5} & \multicolumn{1}{c|}{$T$=4} & \multicolumn{1}{c|}{$T$=4.5} \\
\hline
\bf{\NEAR{} ($M_i$=7)}  &388.42  &365.15  &358.16  &354.62  &353.30  &352.92 \\
\bf{BSPS in \cite{rayliu}}  &668.78  &623.49  &599.42  &592.28  &590.42  &590.08   \\
\bf{All-on}  &1105.15	&1105.15	&1105.15	&1105.15 &1105.15 &1105.15   \\
\hline
\end{tabular}
\vspace{-5mm}
\end{table}

We examine the sum energy for various values of the delay limit $T$.
The results are summarized in Table \ref{tab:res}.  For the
\NEAR{} results in the table, $M_i$ equals 5 and 7, respectively,
for the 7-cell and 19-cell networks. Note that the table does not
include the results of \ALG{} for the 19-cell network, because the
global optimum for this network size is beyond the reach of \ALG{}.
The TDMA scheme (see Section \ref{sec:twosimple}) is not
included since TDMA is infeasible for the delay limits used in
Table \ref{tab:res}.

We make the following observations from the results in
Table~\ref{tab:res}.  First, except for All-on that is insensitive to
$T$ by design, higher QoS requirement (i.e., smaller $T$) requires
higher sum energy. The amount of energy difference is, however,
relatively small for the largest and smallest values of $T$.  Thus
having a larger time limit, or, equivalently, lower QoS requirement,
does not give significant reduction of energy consumption. From the results,
energy saving comes mainly from optimizing cell cluster formation and
activation time duration.

\ALG{} leads to the global optimum and hence the minimum sum energy, 
whereas All-on requires the highest energy consumption by its nature,
as can been seen in the table.  Among the sub-optimal schemes,
\NEAR{} yields the best performance.  Indeed, for the 7-cell network
\NEAR{} consistently achieves less than $1\%$ deviation 
from global optimality.  The BSPS scheme performs rather close to
global optimality for the 7-cell network. For the network with larger
size, however, \NEAR{} leads to significantly better results.

\subsection{Solution Characteristics}

\begin{figure}[htbp]
\vspace{-3mm}
\centering
\includegraphics[scale=0.4]{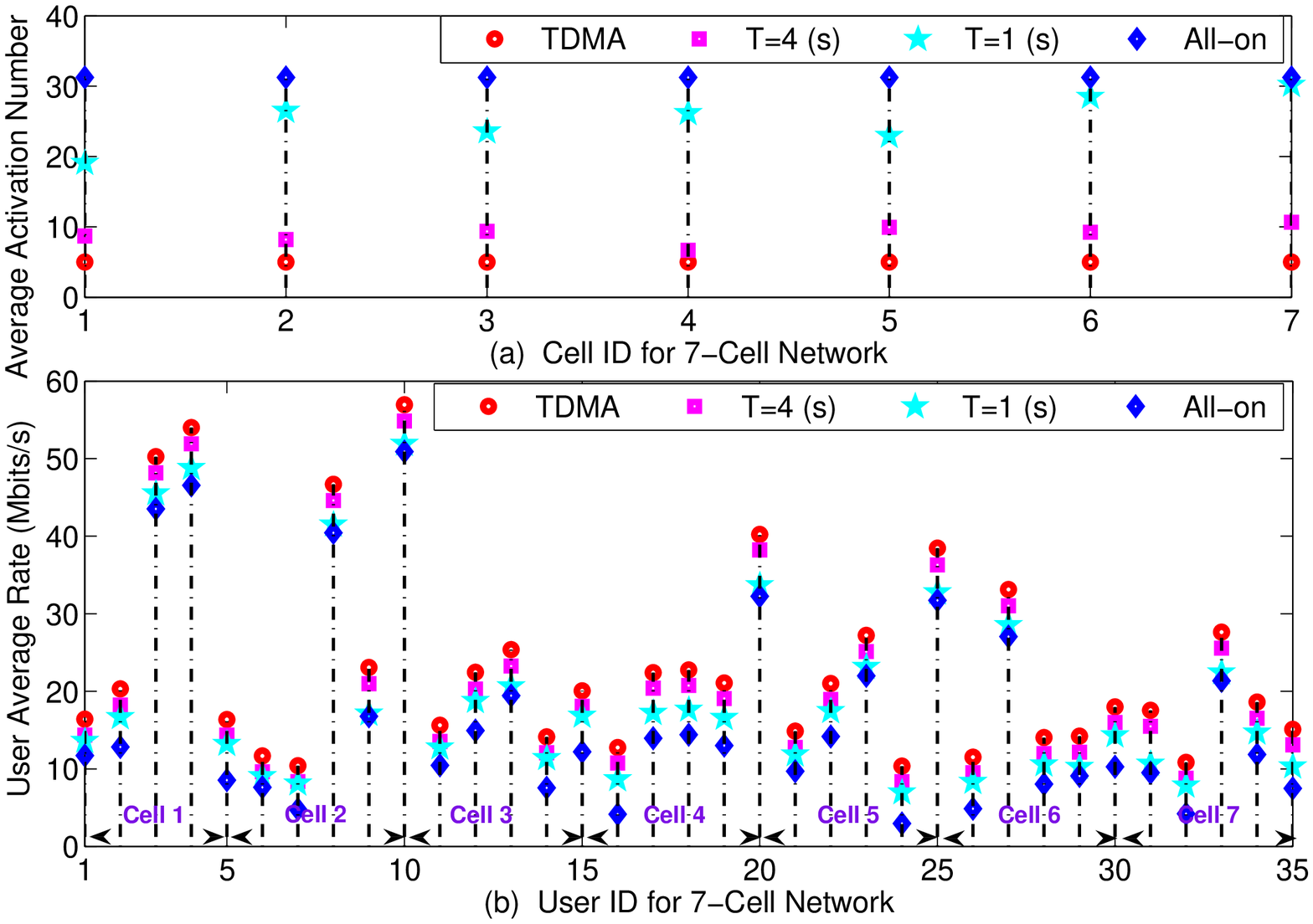}
\vspace{-0.65cm}
\caption{The average number of cell activations and user rate at optimum.}
\label{fig:extra}
\end{figure}

To gain further insights, we consider the
average number of activations of the cells and the average data rate
of the users in TDMA, All-on, and the optimal schedules for $T=1$ and
$T=4$.  The results are displayed in Figure~\ref{fig:extra} for the
7-cell network.

For TDMA, every cell is activated as many times as the number of
users in the cell in Figure~\ref{fig:extra}.  This observation
verifies Lemma~\ref{th:uetdma}, that is, TDMA at the BS level also
implies time-division access of its users.  Because the users are
served one at a time in TDMA, the rate is the highest possible, as can
be seen from Figure~\ref{fig:extra}(b). By Theorem~\ref{th:tdma}, one
would expect that, when the time limit of serving the user demand
becomes more restrictive, the optimal schedule has to use clusters of
larger size, and consequently it is more likely that a BS will appear
in multiple clusters for activation.  This is confirmed by comparing
the results for $T=1$ and $T=4$ in Figure~\ref{fig:extra}(a). Note
that, although the average user rate is lower for small
$T$ in Figure~\ref{fig:extra}(b), the demand can still be served in
shorter time because of the increased number of activations. For All-on,
there is no interruption in transmission, though the user rate
is lowest due to inter-cell interference among all the BSs.  We note
that for All-on, the optimal schedule uses only one cluster of all the
BSs, but the cluster is activated with multiple rate vectors with
optimized activation time durations. From Figure~\ref{fig:extra}(a),
the number of rate vectors used is less than $J+1=36$, which is
consistent with Lemma~\ref{th:lpbasic}.

\subsection{Performance of \BOUND{} in Bounding  Optimum}

\begin{figure}[htbp]
\begin{minipage}[t]{0.45\textwidth}
\centering
\includegraphics[scale=0.345]{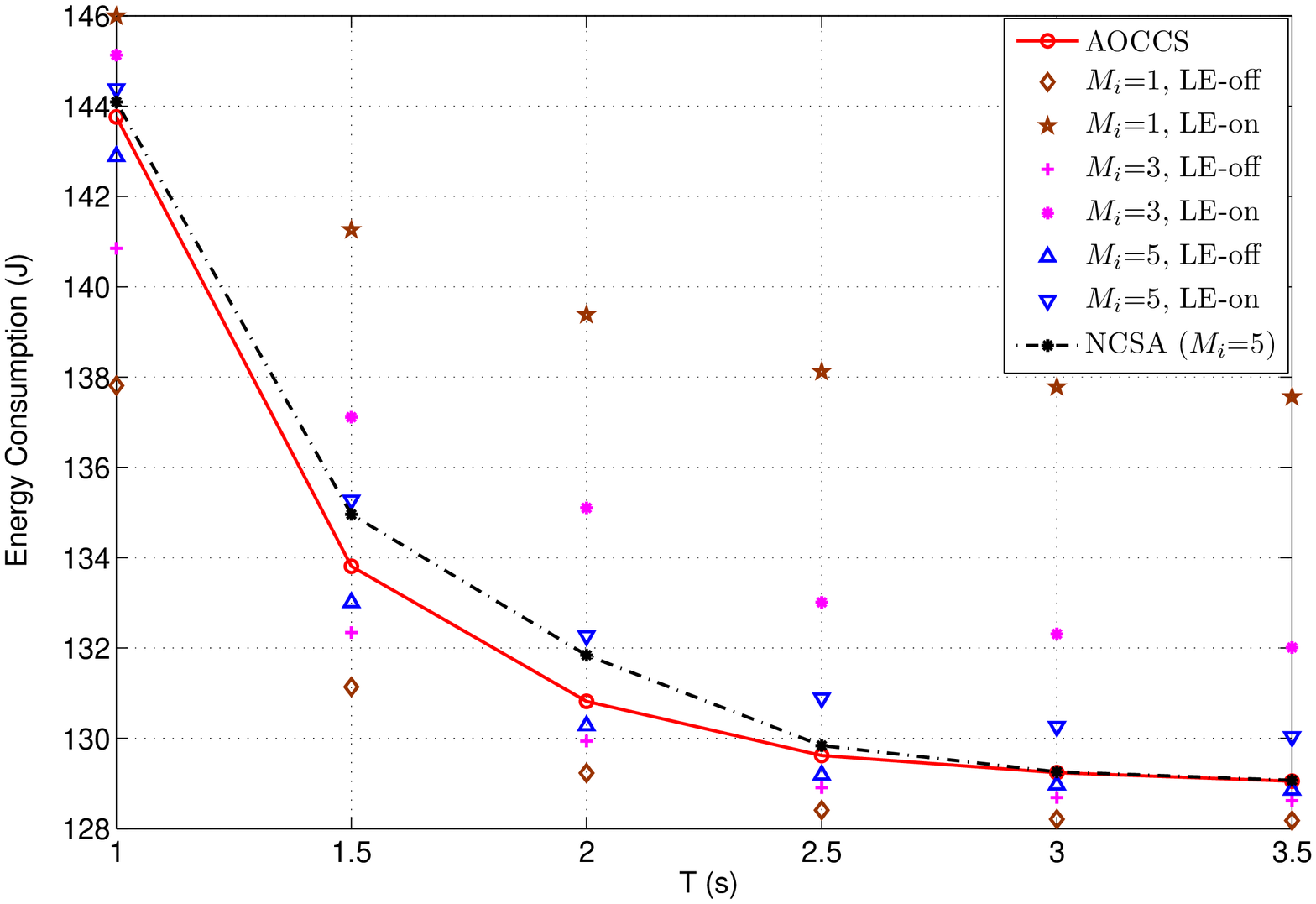}
\vspace{-1.7cm}
\caption{\BOUND{} in bounding optimal solution for the 7-cell network.}
\label{fig:bound7}
\end{minipage}
\hspace{1.0cm}
\begin{minipage}[t]{0.45\textwidth}
\centering
\includegraphics[scale=0.3]{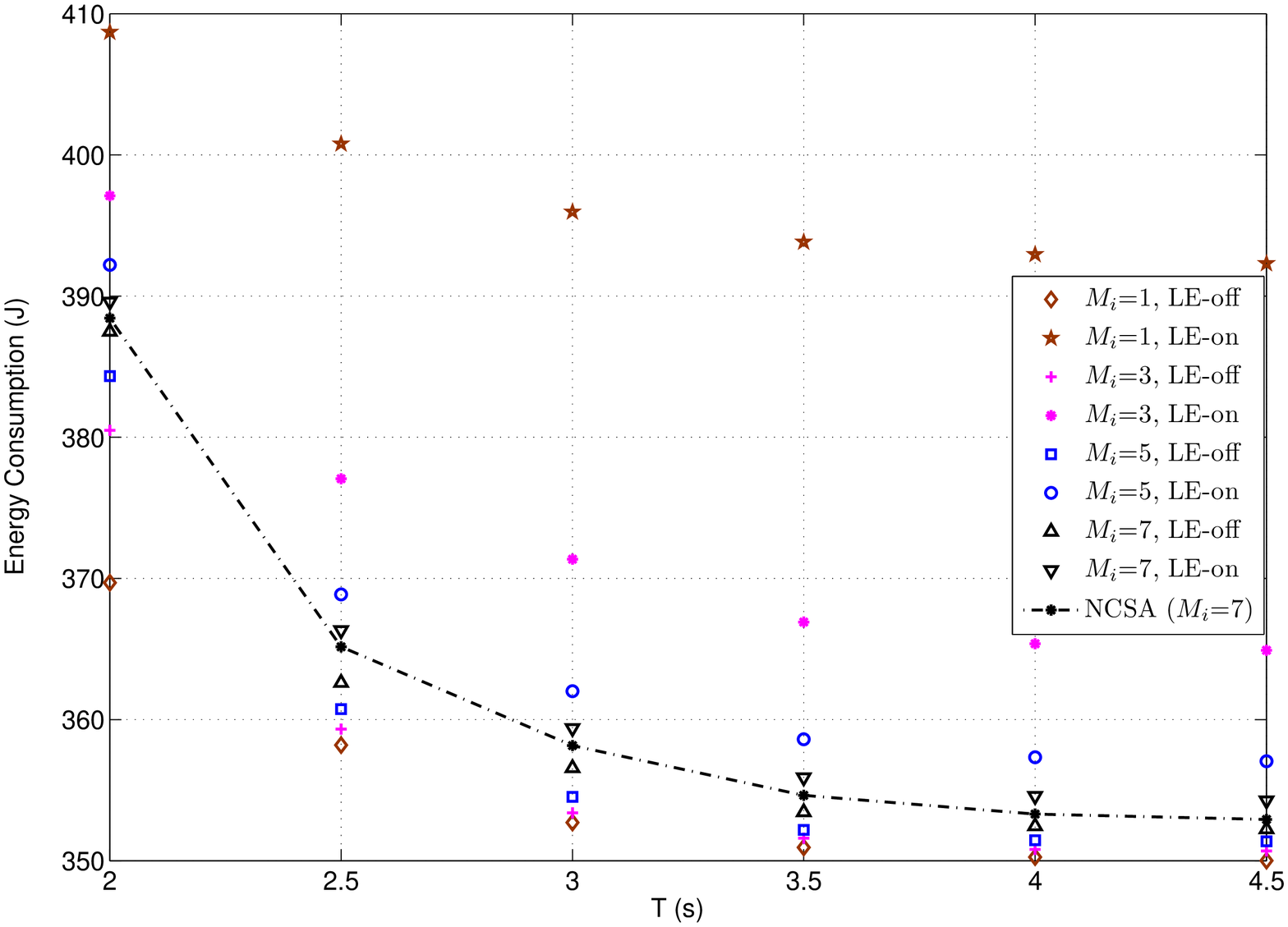}
\vspace{-1.7cm}
\caption{\BOUND{} in bounding  optimal solution for the 19-cell network.}
\label{fig:bound19}
\end{minipage}
\vspace{-1.0cm}
\end{figure}

We examine the accuracy of the estimation of global optimum via
\BOUND{}, and set this in perspective to \ALG{} and \NEAR{}.  The
results, given as sum energy versus delay limit $T$, are shown in
Figures \ref{fig:bound7} and
\ref{fig:bound19}. For a comprehensive
performance picture, $M_i$ is successively increased in the two
figures. For each value of $M_i$, a pair of markers is used to show
the upper and lower bounds of the global optimum.  The gaps between
the upper and lower bounds from \BOUND{}, averaged over $T$ for
selected values of $M_i$, are further detailed in Table~\ref{tab:gap}.  In
addition to setting $M_i$ uniformly for all cells, Table~\ref{tab:gap}
also contains results of 
setting $M_i$ to be the number of cell $i$'s one-hop neighbor cells.
For example, in the 7-cell network, $M_i = 6$ for the center
cell and $M_i = 3$ for the other cells.  The results obtained with
this setting is referred to as ``Neighbor-$M_i$''.

\vspace{-0.5cm}


From the two figures and Table~\ref{tab:gap}, augmenting the size of
local enumeration of interference (i.e., parameter $M_i$)
leads to progressively tighter bounding intervals. Note that, even
with $M_i$ being as small as one, that is, only a single
neighboring BS is accounted for, the accuracy remains
satisfactory -- the relative difference of the upper and lower
bounds of global optimum is less than $8\%$ and $12\%$, respectively,
for the two networks. We observe that when $T$ increases, the lower
bound from LE-off tends to improve in relation to \ALG{} or \NEAR{},
whereas the upper bound from LE-on does not. This is because LE-on
over-estimates interference, and for large $T$ the error grows because
optimal clusters tend to be small (cf.~Theorem \ref{th:tdma}).
For LE-off, increasing $T$ has the reverse effect.

\begin{table}[htbp]
\vspace{-5mm}
\addtolength{\tabcolsep}{-5pt}
\caption {Average accuracy of the bounding interval from \BOUND{}.}
\vspace{-7mm}
\label{tab:gap}
\centering
\begin{tabular}[t]{|c|c|c|}
\cline{2-3}
\multicolumn{1}{c}{}& \multicolumn{2}{|c|}{\T Relative difference between $E^*_{\text{LE-on}}$ and $E^*_{\text{LE-off}}$, \B}\\
\multicolumn{1}{c}{}& \multicolumn{2}{|c|}{\T $(E^*_{\text{LE-on}}-E^*_{\text{LE-off}})/E^*_{\text{LE-off}}\times100\%$ \B}\\
\cline{1-3}
\backslashbox{$M_i$}{Network} & \multicolumn{1}{c|}{ \ \ \ \bf{7-cell Network} \ \ \ } & \multicolumn{1}{c|}{\bf{19-cell Network}} \\
\cline{2-3}
\cline{1-3}
$\bm{M_i=1}$ &7.31\% & 11.87\% \\
$\bm{M_i=3}$ &3.21\% &4.49\% \\
$\bm{M_i=5}$ & 1.25\% &1.92\% \\
$\bm{M_i=7}$ &0\% & 0.71\% \\
$\bf Neighbor$-$\bm{M_i}$ &0.58\% &0.98\% \\
\hline
\end{tabular}
\vspace{-5mm}
\end{table}

\NEAR{} combines LE-on with post-processing.  From Figure \ref{fig:bound7},
\NEAR{} performs extremely close to global optimum for the 7-cell
network -- the relative deviation is merely 0.7\% or less.  For the
19-cell network, global optimum is not available for evaluating
\NEAR{}. However, the lower bound of global optimum, 
derived from LE-off, reveals that the deviation
from global optimum is within 1\%. This demonstrates the performance
of \NEAR{} as well as the usefulness of the bounding scheme.
Moreover, from the last row of Table~\ref{tab:gap}, setting $M_i$
based on the number of one-hop neighbors significantly outperforms
uniformly setting $M_i=5$, while the problem sizes in \BOUND{} are
comparable for the two settings. 
The cell-adaptive choice of $M_i$ achieves similar performance as
setting $M_i =7$. However, the 
problem size is considerably smaller in the former because 
$M_i<7$ for most cells.

\section{Conclusions}
\label{sec:conclusions}

We have considered optimal base station clustering and scheduling with
the objective of minimizing energy consumption. Theoretical insights
and mathematical formulations have been provided. For problem
solution, we have presented a column generation approach, as well as a
local enumeration scheme. The latter effectively addresses the
difficulty of optimal cluster formation that is of combinatorial
nature. Integrating column generation with local enumeration not only
leads to flexibility in balancing optimality with scalability, but
also yields lower and upper bounds confining the global
optimum. Numerical results demonstrate that the algorithmic notions
result in significant improvement in energy saving in comparison to
existing schemes. In addition, the BS clustering and scheduling
solutions that have been obtained are very close to global optimum.

The work in this paper provides a theoretical framework of
optimizing BS clustering and activation. The
proposed framework can be potentially implemented using the almost
blank subframes (ABS) scheme defined in 3GPP Release 10.  The BSs
during their deactivation time durations
can be set to the ABS mode, in which only control
channels can be used with very low power, whereas the active BSs are
in normal transmission mode.  Also, from a scalability standpoint, the
use of \NEAR{} with local enumeration of interference has two
implications. First, the problem size grows only linearly instead of
exponentially in the number of BSs. Second, performance calculation
for each BS needs to consider the neighboring BSs only. As such, the
signaling cost for implementing the framework is reasonable.

An extension of the current work is to investigate the potential of
power control.  Base station clustering with cooperative multi-point
transmission is another topic for future studies.




\end{document}